\documentclass[prb,twocolumn,showpacs,amsmath,amssymb]{revtex4}
\pdfoutput=1

\usepackage{graphicx}
\usepackage{dcolumn}
\usepackage{bm}

\def\s{\sigma}
\def\pr{^{\prime}}
\def\f{\frac}
\def\LB{\left(}
\def\RB{\right)}
\def\u{\uparrow}
\def\d{\downarrow}
\def\ga{\gamma}

\begin{document}


\title{The Kohn-Sham system in one-matrix functional theory}

\author{Ryan Requist}
 \email{Ryan.Requist@physik.uni-erlangen.de}
\author{Oleg Pankratov}%
\affiliation{%
Lehrstuhl f\"ur theoretische Festk\"orperphysik \\
Friedrich-Alexander-Universit\"at Erlangen-N\"urnberg \\
}%

\date{\today}

\begin{abstract}

A system of electrons in a local or nonlocal external potential can be 
studied with 1-matrix functional theory (1MFT), which is similar to 
density functional theory (DFT) but takes the one-particle reduced 
density matrix (1-matrix) instead of the density as its basic variable. 
Within 1MFT, Gilbert derived [PRB \textbf{12}, 2111 (1975)] effective 
single-particle equations analogous to the Kohn-Sham (KS) equations in 
DFT. The self-consistent solution of these 1MFT-KS equations  
reproduces not only the density of the original electron system
but also its 1-matrix.  While in DFT it is usually possible to reproduce the 
density using KS orbitals with integer (0 or 1) occupancy, in 1MFT
reproducing the 1-matrix requires in general fractional occupancies. 
The variational principle implies that the KS eigenvalues of all 
fractionally occupied orbitals must collapse at self-consistency to a single level,
equal to the chemical potential.
We show that as a consequence of the degeneracy the iteration of the 
KS equations is intrinsically divergent.  Fortunately, the level shifting 
method, commonly introduced in Hartree-Fock calculations, is always able 
to force convergence.  We introduce an alternative derivation of the 1MFT-KS 
equations that allows control of the eigenvalue collapse by constraining the 
occupancies. As an explicit example, we apply the 1MFT-KS scheme 
to calculate the ground state 1-matrix of an exactly solvable two-site 
Hubbard model. 
\end{abstract}

\maketitle

\section{\label{sec:introduction} Introduction}

Density functional theory (DFT) benefits from operating with
the electron density, which as a function of just three coordinates is
much easier to work with than the full many-body wavefunction.  According 
to the Hohenberg-Kohn (HK) theorem, \cite{hohenberg1964} the density of 
an electron system in a local external potential $v(\vec{r})$ may 
be found by minimizing a universal energy functional $E_v[n]$, whose basic 
variable is the density.  Remarkably, the density uniquely determines the 
ground state wavefunction (if it is nondegenerate), i.e., there can be 
only one ground state wavefunction yielding a given density, no matter 
what $v(\vec{r})$ is.  However, if the external potential is nonlocal, then the 
density alone is generally not sufficient to uniquely determine the ground 
state (see Appendix A for a simple example). Gilbert \cite{gilbert1975} 
extended the HK theorem to systems with nonlocal and 
spin dependent external potential $v(x,x\pr)$, where $x=(\vec{r},\s)$.  It was 
proved that i) the ground state wavefunction is uniquely determined by the 
ground state 1-matrix (one-particle reduced density matrix) and ii) there 
is a universal energy functional $E_v[\ga]$ of the 1-matrix, which attains 
its minimum at the ground state 1-matrix.  The 1-matrix is defined as
\begin{equation}
\gamma(x,x\pr) = N \int dx_2 \ldots dx_N  \rho(x, x_2, \ldots x_N ; x\pr,x_2, \ldots x_N), 
\label{eqn:1-matrix}
\end{equation}
where $\int dx = \sum_{\s} \int d^3r$ and 
$\hat{\rho} = \sum_i w_i \left| \Psi_i \right> \left< \Psi_i \right|$
is the full $N$-electron density matrix with ensemble weights $w_i$ such 
that $\sum_i w_i = 1$.  An external potential may be nonlocal with respect 
to the space coordinates and/or the spin coordinates.  For example, 
pseudopotentials are nonlocal in space, and Zeeman coupling 
$-(\hbar|e|/mc)\vec{B}\cdot\vec{\sigma}$, where $\vec{\sigma}$ is the 
vector of Pauli matrices, is nonlocal in spin space. The coupling of 
electron motion and an external vector potential, 
$(|e|/2mc) (\vec{p}\cdot\vec{A} + \vec{A}\cdot\vec{p})$, 
may also be treated 
as a nonlocal potential because $\vec{p}$ is a differential operator.  
It is rather intuitive that for such external potentials, which couple to 
the system in more complex ways than the local potential $v(\vec{r})$, it is 
necessary --- in order to permit statements analogous to the HK theorem 
--- to refine the basic variable accordingly.  Hence spin-DFT, 
\cite{vonbarth1972,gunnarsson1976} whose basic variables are the density 
and the magnetization density, applies to systems with Zeeman coupling.  
Current-DFT, \cite{vignale1987,vignale1988} whose basic variables are the 
density and the paramagnetic current density, has the scope to treat systems 
in which the current is coupled to an external magnetic field.  Generally, 
if one considers an external potential that is nonlocal in space and spin, 
the necessary basic variable is the one-matrix,\cite{gilbert1975} which 
contains all of the single-particle information of the system, including 
the density, magnetization density and paramagnetic current density.

The DFT-type approach that takes the 1-matrix as basic variable will be 
referred to here as 1-matrix functional theory (1MFT).  As in DFT, an 
exact and explicit energy functional is generally unknown. An important 
difference between 1MFT and DFT is that the kinetic energy is a simple 
linear functional of the 1-matrix, while it is not a known functional of 
the density.  Thus, in 1MFT the only part of the energy not known 
explicitly is the electron-electron interaction energy $W[\gamma]$.  
Several approximate 1-matrix energy functionals have been proposed and 
tested recently (see Refs.~\onlinecite{gritsenko2005} and 
\onlinecite{lathiotakis2007} and references therein.)  Notably, the 
so-called BBC$n$ approximations,\cite{gritsenko2005} which are modifications of the 
Buijse-Baerends functional,\cite{buijse2002} have given fairly accurate results 
for the potential energy curves of diatomic molecules\cite{gritsenko2005} 
and the momentum distribution and correlation energy of the homogeneous 
electron gas.\cite{lathiotakis2007}  In Ref.~\onlinecite{lathiotakis2007}, 
a density dependent fitting parameter was introduced into the BBC1 functional
such that the resulting functional yields the correct correlation energy of
the homogeneous electron gas at all values of density.
There is also the prospect of using 1MFT to obtain accurate estimates 
for the band gaps of non-highly correlated insulators.\cite{helbig2007}  
Many of the approximate functionals that have been proposed are similar to 
an early approximation by M\"uller.\cite{mueller1984}  

Actual calculations in 1MFT are more difficult than in DFT.  The 
energy functional $E_v[\gamma]$ must be minimized in a space of higher dimension 
because the 1-matrix is a more complex quantity than the density.  In the calculations 
cited above, the energy has been minimized directly by standard methods, e.g., 
the conjugate gradient method.  In DFT the energy is generally not 
minimized by such direct methods.  Instead, the Kohn-Sham (KS) scheme\cite{kohn1965} 
provides an efficient way to find the ground state density. 
In this scheme, one introduces an auxiliary system of $N$ 
noninteracting electrons, called the KS system, which experiences an 
effective local potential $v_s(\vec{r})$.  This effective potential is a 
functional of the density such that the self-consistent ground state of 
the KS system reproduces the ground state density of the interacting 
system.  It is interesting to ask whether there is also a KS scheme in 
1MFT.  The question may be stated as follows:  does there exist a 
1-matrix dependent effective potential $v_s(x,x\pr)$ such that, at 
self-consistency, a system of noninteracting electrons experiencing this 
potential reproduces the exact ground state 1-matrix of the interacting 
system?  Although Gilbert derived such an effective potential,\cite{gilbert1975} 
the implications were thought to be ``paradoxical'' because the KS 
system was found to have a high (probably infinite) 
degree of degeneracy.  Evidently, the KS eigenvalues in 1MFT do not 
have the meaning of approximate single-particle energy levels, in 
contrast to DFT and other self-consistent-field theories, where the 
eigenvalues may often be interpreted as the negative of ionization 
energies, owing to Koopmans' theorem.  
The status of the KS scheme in 1MFT appears to have 
remained unresolved, \cite{valone1980,tal1985} and recently it has been 
argued that the KS scheme does not exist in 1MFT.
\cite{schindlmayr1995,lathiotakis2005,helbig2007,lathiotakis2007}  
Gilbert derived the KS equations from the stationary principle for the 
energy.  The KS potential was found to be 
\begin{equation}
 v_s(x,x\pr) = v(x,x\pr) + \delta W/\delta \gamma(x\pr,x). 
 \label{eqn:KSpot}
\end{equation}
In this article, we propose an alternative derivation of the KS equations, 
which, in our view, gives insight into the nature of the 
``paradoxical'' degeneracy of the KS system.

One-matrix energy functionals are often expressed in terms of the 
so-called natural orbitals and occupation numbers. \cite{loewdin1955}  
This makes them similar to ``orbital dependent'' functionals in DFT.  
The natural orbitals are the eigenfunctions of the 1-matrix, and the 
occupation numbers are the corresponding eigenvalues.\cite{loewdin1955}
These quantities play a central role in 1MFT.  Recently, it was shown 
that when a given energy functional is expressed in terms of the 
natural orbitals and occupation numbers, the KS potential can be found 
by using a chain rule to evaluate the functional derivative in 
Eq.~\ref{eqn:KSpot}.\cite{pernal2005} 

Although the concept of the KS system can indeed be extended to 1MFT, it has 
in this setting some very unusual properties. 
In particular, the KS orbitals must be fractionally occupied, 
for otherwise the KS system could not reproduce the 1-matrix of the interacting
system, which always has noninteger eigenvalues (occupation numbers).  
This is different from the situation in DFT, where it is usually possible to reproduce 
the density using only integer (0 or 1) occupation numbers, or in any case, only a 
finite number of fractionally occupied states. 
Due to the necessity of fractional occupation numbers, the 1MFT-KS system 
cannot be described by a single Slater determinant.  However, we find that it can be 
described by an ensemble of Slater determinants, i.e., a mixed state.  In order that 
the variational principle is not violated, all the states that comprise the ensemble 
must be degenerate.  This implies that the eigenvalues of all fractionally occupied 
orbitals collapse to a single level, equal to the chemical potential.  
The degeneracy has important consequences for the solution 
of the KS equations by iteration.  We prove that the iteration of the KS equations 
is intrinsically divergent because the KS system has a divergent response
function $\chi_s=\delta \ga/\delta v_s$ at the ground state.  Fortunately, 
convergence can always be obtained with the level shifting method.\cite{saunders1973} 
To illustrate explicitly the unique properties of 
the 1MFT-KS system, we apply it to a simple Hubbard model with two sites.  The 
model describes approximately systems which have two localized orbitals with a 
strong on-site interaction, e.g., the hydrogen molecule 
with large internuclear separation. \cite{aryasetiawan2002} The Schr\"odinger
equation for this model is exactly solvable, and we find that the KS equations 
in 1MFT and in DFT can be derived analytically.  It is interesting to compare 
1MFT and DFT in this context.  We demonstrate that divergent behavior will appear 
also in DFT when the operator $1- \chi_s \chi^{-1}$, where $\chi$ and
$\chi_s$ are the density response functions of the interacting and KS systems, 
respectively, has any eigenvalue with modulus greater than 1.  In this expression 
the null space of $\chi$ is assumed to be excluded. 

This article is organized as follows.  In Sec.~\ref{sec:KS}, we derive the KS 
equations in 1MFT and discuss how to solve them self-consistently by
iteration. In Sec.~\ref{sec:Hubbard}, we compare three approaches to ground state 
quantum mechanics --- direct solution of the Schr\"odinger equation, 
1MFT and DFT --- by using them to solve the two-site Hubbard model.

\section{\label{sec:KS} Kohn-Sham system in 1MFT}

It is not obvious that a KS-type scheme exists in 1MFT for the 
following reason.  Recall that in DFT the KS system consists of
$N$ noninteracting particles and reproduces the density of the 
interacting system.  The density of the KS system, if it is
nondegenerate, is the sum of contributions of the $N$ lowest 
energy occupied orbitals
\begin{equation}
 n(\vec{r}) = \sum_{i}^{occ} \left| \phi_i(\vec{r}) \right|^2.  
 \label{eqn:density}
\end{equation}
On the other hand, in 1MFT the KS system should reproduce the
1-matrix of the interacting system.  The eigenfunctions of the 
1-matrix are the so-called natural orbitals, and the eigenvalues
are the corresponding occupation numbers. \cite{loewdin1955} 
Occupying the $N$ lowest energy orbitals in analogy to 
(\ref{eqn:density}), one obtains
\begin{equation}
 \gamma(x,x\pr) = \sum_i^{occ} \phi_i(x) \phi_i^*(x\pr).
 \label{eqn:1-matrix:nondeg}
\end{equation}
Such an expression, in which the orbitals have only integer 
(0 or 1) occupation, cannot reproduce the 1-matrix of 
an interacting system because the orbitals of an 
interacting system have generally fractional occupation 
(see the discussion in the following section.)  The 
difference between the 1-matrix in (\ref{eqn:1-matrix:nondeg})
and the 1-matrix of an interacting system is  
clearly demonstrated by the so-called idempotency property.
The 1-matrix in (\ref{eqn:1-matrix:nondeg}) is idempotent, i.e.,
$\int dx\pr \gamma(x,x\pr) \gamma(x\pr,x^{\prime \prime})=\gamma(x,x^{\prime\prime})$, 
while the 1-matrix of an interacting system is  
never idempotent.  However, if the KS system is degenerate 
and its ground state is an ensemble state, the 1-matrix 
becomes
\begin{equation}
 \gamma(x,x\pr) = \sum_i f_i \phi_i(x) \phi_i^*(x\pr)
 \label{eqn:1-matrix:diag}
\end{equation}
with fractional occupation numbers $f_i$.  The $N$-particle 
ground state density matrix of the KS system is 
$\hat{\rho}_s = \sum_i w_i \left|\Phi_i\right> \left< \Phi_i\right|$,
where the $\Phi_i$ are Slater determinants each formed from 
$N$ degenerate KS orbitals.  The occupation numbers $f_i$ 
are related to the ensemble weights $w_i$ by
\begin{equation}
 f_i = \sum_j w_j \Theta_{ji},
\end{equation}
where $\Theta_{ji}$ equals $1$ if $\phi_i$ is one of the orbitals in 
the determinant $\Phi_j$ and 0 otherwise. \cite{gross1990}

\subsection{\label{ssec:defn} Derivation of the 1MFT Kohn-Sham equations}

In this section, we discuss Gilbert's derivation\cite{gilbert1975} 
of the KS equations in 1MFT and propose an alternative derivation.
We begin by reviewing the definition of the universal 1-matrix
energy functional $E_v[\ga]$.

One-matrix functional theory describes the ground state of a 
system of $N$ electrons with the Hamiltonian 
$\hat{H} = \sum_{i=1}^N (\hat{t}_i+\hat{v}_i) + \hat{W}$, where 
$\hat{t}=-\nabla_r^2/2$ is the kinetic energy operator, $\hat{v}$ 
is the local or nonlocal external potential operator, and 
$\hat{W}=\sum_{i<j} |\vec{r}_i-\vec{r}_j|^{-1}$ is the electron-electron 
interaction (in atomic units $\hbar=m=e=1$).  The ground state 
1-matrix and ground state energy can be found by minimizing the 
functional
\begin{equation}
 E_v[\ga] = Tr((\hat{t}+\hat{v})\ga) + W[\ga], \label{eqn:HK}
\end{equation}
where
\begin{equation}
 W[\ga] = \big< \Psi_0 \big| \hat{W} \big| \Psi_0 \big>.
 \label{eqn:W:gs}
\end{equation}
By extending the HK theorem, Gilbert proved\cite{gilbert1975} that a nondegenerate 
ground state wavefunction, $\Psi_0$, is uniquely determined by 
the ground state 1-matrix, i.e., $\Psi_0$ is a functional of 
$\ga$. For this reason the interaction energy, as defined in 
(\ref{eqn:W:gs}), is a functional of $\ga$.  It is apparent that
(\ref{eqn:W:gs}) defines $W[\ga]$ only for $\ga$ that are the 
ground state 1-matrices of some system (with Hamiltonian $\hat{H}$).  
In this article, a 1-matrix is said to be $v$-representable (VR) if 
it is the ground state 1-matrix of some system with local \emph{or nonlocal} 
external potential. 
Gilbert remarked (see the discussion between equations (2.24) and (2.25) 
in Ref.~\onlinecite{gilbert1975}) that, in principle, the domain 
of $W[\gamma]$ can be extended to the space of ensemble $N$-representable (ENR)
1-matrices.  A 1-matrix is said to be ENR if it can be constructed
via (\ref{eqn:1-matrix}) from some $N$-particle density matrix 
$\hat{\rho}=\sum_i w_i \left| \Psi_i \right> \left< \Psi_i \right|$,
which is not required to be a ground state ensemble. 
One possible extension to the ENR space is provided by the 
so-called constrained search functional\cite{levy1979,valone1980}
\begin{equation}
 W[\ga] = \textrm{min}_{\hat{\rho} \rightarrow \ga} Tr(\hat{W} \hat{\rho}),
 \label{eqn:levy}
\end{equation}
where the interaction energy $Tr(\hat{W} \hat{\rho})$ is minimized 
in the space of $N$-particle density matrices $\hat{\rho}$ 
that yield $\ga$ via (\ref{eqn:1-matrix}).  The definition (\ref{eqn:levy}) 
is a natural extension to the ENR space because when it is adopted 
(\ref{eqn:HK}) may be expressed as
\begin{equation}
 E_v[\ga] = \textrm{min}_{\hat{\rho}\rightarrow \ga} Tr(\hat{H}\hat{\rho}).
 \label{eqn:HK2}
\end{equation}
This is a variational functional which attains its minimum at the
ground state 1-matrix, as seen from 
\begin{equation}
 \textrm{min}_{\ga} E_v[\ga] = \textrm{min}_{\hat{\rho}} Tr(\hat{H}\hat{\rho}) = E_0,
 \label{eqn:minimum}
\end{equation}
where $E_0$ is the ground state energy.
The extension to the ENR domain is significant, especially for 
applications of the variational principle, because the conditions 
a 1-matrix must satisfy to be ENR are known and simple to impose 
on a trial 1-matrix, while the conditions for $v$-representability 
are unknown in general.  The necessary and sufficient conditions
\cite{coleman1963} for a 1-matrix $\gamma$ to be ENR are 
i) $\gamma$ must be Hermitean, ii) $\int dx \gamma(x,x) = N$, and 
iii) all eigenvalues of $\gamma$ (occupation numbers) must lie in 
the interval $[0,1]$.  The third condition is a consequence of the
Pauli exclusion principle.

The 1MFT-KS equations were derived\cite{gilbert1975} from the 
stationary conditions for the energy with respect to arbitrary 
independent variations of the natural orbitals $\phi_i$ and 
angle variables $\theta_i$ chosen to parametrize the occupation 
numbers according to $f_i=\cos^2 \theta_i$ ($0\leq \theta_i \leq \pi/2$). 
For the purpose of describing a variation in the ENR space, this 
set of variables, namely $\{\delta \phi_i,\delta \phi_i^*,\delta \theta_i\}$,
is redundant.  An arbitrary set of such variations may or may not 
correspond to an ENR variation, and when it does the variations 
will not be linearly independent. This causes no difficulty, of 
course, because it is always possible to formulate stationary 
conditions in a space whose dimension is higher than necessary,
provided the appropriate constraints are enforced with Lagrange 
multipliers.  Accordingly, the Lagrange multiplier terms
$\sum_{ij} \lambda_{ij} (\left<\phi_i|\phi_j\right>-\delta_{ij})$
which maintain the orthogonality of the orbitals and the term
$\mu (\sum_i f_i - N)$ which maintains the total particle number 
were introduced. The KS equations were found to be
\begin{equation}
 (\hat{t}+\hat{v}_s) \left|\phi_i \right> = \epsilon_i \left|\phi_i \right>,
\end{equation}
where the kernel of the effective potential is
$v_s(x,x\pr) = v(x,x\pr)+ \delta W/\delta \gamma(x\pr,x)$,
if the functional derivative exists. The stationary conditions 
imply that all fractionally occupied KS orbitals have the same 
eigenvalue $\epsilon_i = \mu$. Gilbert described this result as 
``paradoxical'' because in interacting systems essentially all 
orbitals are fractionally occupied.

The above stationary conditions assume $E_v$ to be stationary
with respect to variations in the ENR space (except variations 
of occupation numbers equal to exactly 0 or 1 in the ground 
state, which are excluded by the parametrization). However, it 
is not known, in general, whether $E_v$ is stationary in the ENR 
space; the minimum property (\ref{eqn:minimum}) ensures only that 
it is variational. Recall that 
the ENR space consists of all $\ga$ that can be constructed from an 
ensemble, and the energy of an \emph{ensemble} is not stationary 
with respect to variations of the many-body density matrix 
$\hat{\rho}$.  Therefore, the stationary conditions applied in 
the ENR space may be too strong.  
On the other hand, the quantum mechanical variational 
principle guarantees that $E_v$ is stationary with respect to 
variations in the VR space,\footnote{\label{foot:thm} The following 
theorem\cite{gelfand1963} implies that the functional $E_v[\ga]$ 
is stationary, i.e., $\delta E_v = 0$ with respect to all 
variations in the VR space. Theorem -- A necessary condition for 
the differentiable functional $J[y]$ to have an extremum (minimum) 
for $y=y_0$ is that its first variation vanish for $y=y_0$, i.e., 
that $\delta J[h] = 0$ for $y=y_0$ and all admissible variations 
$h$.  Thus, the functional $E_v[\ga]$ must be stationary because 
it is differentiable in the VR space and the quantum mechanical
variation principle (Rayleigh-Ritz variational principle) implies 
that it is minimum at the ground state 1-matrix $\ga=\ga_{gs}$. 
The differentiability of $E_v[\ga]$ follows from the 
differentiability of the mapping $\gamma \rightarrow \Psi_0$ in 
the VR space, if the ground state $\Psi_0$ is nondegenerate.}
but it is not known how to determine whether a given $\ga$ is VR.  
Hence, it is not known how to constrain the variations of $\ga$ 
to the VR space.  Nevertheless, it may be that in some systems 
the entire neighborhood of $\ga_{gs}$ in the ENR space is also 
VR.  In such cases, $E_v$ is stationary in the ENR space, and 
the stationary conditions applied in Ref.~\onlinecite{gilbert1975} 
are satisfied at the ground state.


We find it helpful to construct an alternative derivation of the 
KS equations.  Consider the energy functional
\begin{equation}
 \mathcal{G}_v[\ga] = E_v[\ga] - \sum_j \epsilon_j (f_j-q_j), \label{eqn:total}
\end{equation}
where $\epsilon_j=\epsilon_j(\{q_k\})$ are Lagrange multipliers 
that constrain the occupation numbers $f_j$ of the natural orbitals 
to chosen values $q_j$, which satisfy $0\leq q_j \leq 1$ and
$\sum_j q_j=N$.  These Lagrange multipliers allow us to investigate 
the degeneracy of the KS eigenvalues, which leads to the ``paradox'' 
described by Gilbert.  We have omitted the Lagrange multipliers 
$\lambda_{ij}$ used in Gilbert's derivation.  They are not necessary 
in our derivation because we formulate the stationary conditions 
with respect to variations of the 1-matrix instead of the orbitals.  
We adopt the definition (\ref{eqn:levy}) for $W[\ga]$, so the domain 
of $\mathcal{G}_v[\ga]$ is the ENR space.  A variation $\delta \ga$ 
will be said to be admissible if $\ga_{gs} + \delta \ga$ is ENR. 
For convenience we assume that the static response function 
$\chi=\delta \ga/\delta v$ for the interacting system under 
consideration has no null vectors apart from the null vectors 
associated with a) a constant shift of the potential (which is a 
null vector also in DFT) and b) integer occupied orbitals, i.e., 
orbitals with occupation numbers exactly 0 or 1.  If $\chi$ has 
additional null vectors, the following derivation must be modified; 
the necessary modifications are discussed below.
Granting the above assumption, $\mathcal{G}_v$ 
is guaranteed to be stationary, and the KS equations can be derived 
from the stationary condition $\delta \mathcal{G}_v = 0$ with 
respect to an arbitrary admissible variation of $\ga$.  The first 
variation of $\mathcal{G}_v$ is 
\begin{eqnarray}
 \delta \mathcal{G}_v &=& Tr((\hat{t}+\hat{v}) \delta \hat{\ga}) + Tr(\hat{w} \delta \hat{\ga})
 - \sum_j \epsilon_j \delta f_j \nonumber \\
 &=& \sum_{ij} \left< \phi_i \left| (\hat{t}+\hat{v}+\hat{w}) \right| \phi_j \right> 
 \big< \phi_j \big| \delta \hat{\ga} \big| \phi_i \big> \nonumber \\
 && - \sum_{ij} \epsilon_j \left<\phi_i | \phi_j \right> 
 \big< \phi_j \big| \delta \hat{\ga} \big| \phi_i \big> \nonumber \\
 &=& \sum_{ij} (h_{ij}- \epsilon_j \delta_{ij} )\:\delta \ga_{ji}, \label{eqn:stat}
\end{eqnarray}
where the variation of the 1-matrix is expressed as 
$\delta \ga_{ij} = \big< \phi_i \big| \delta \hat{\ga} \big| \phi_j \big>$
in the basis of the ground state natural orbitals, and the relation 
$\delta W = Tr(\hat{w} \delta \hat{\ga})$ defines 
a single-particle operator $\hat{w}$. 
In (\ref{eqn:stat}), we have also introduced the definition of the
Hermitean operator
\begin{equation}
 \hat{h}=\hat{t}+\hat{v}+\hat{w},
 \label{eqn:1MFT-ham}
\end{equation}
which will be seen to be the KS Hamiltonian. If the last line of 
(\ref{eqn:stat}) is to be zero for an arbitrary Hermitian matrix $\delta \ga$, 
then we must have $h_{ij}-\epsilon_j \delta_{ij}=0$ for all $i$ and $j$.
ENR condition (i) has been maintained explicitly by requiring the variation
to be Hermitian.  ENR conditions (ii) and (iii) do not impose any constraint 
on the space of admissible variations as they are maintained by the Lagrange 
multipliers $\epsilon_i$.  The matrix elements $h_{ij}$ are functionals of the 1-matrix, 
and the 1-matrix that satisfies the stationary conditions 
$h_{ij}-\epsilon_j \delta_{ij}=0$ can be found by solving self-consistently 
the single-particle equations 
\begin{equation}
 \hat{h} \big| \phi_i \big> = \epsilon_i \big| \phi_i \big>
 \label{eqn:1MFT-KS}
\end{equation}
together with (\ref{eqn:1-matrix:diag}).
These are the KS equations in 1MFT. If they are solved self-consistently with 
the occupation numbers fixed to the values $q_i$, they give the
orbitals which minimize $E_v$ subject to the constraints $f_i=q_i$.
The KS potential is $\hat{v}_s = \hat{v}+\hat{w}$.  The term 
$\hat{w}$ is the effective contribution of the electron-electron 
interaction to the KS potential.  In coordinate space, its kernel is
\begin{eqnarray}
 w(x,x^{\prime}) &=& \big< x \big| \hat{w} \big| x^{\prime} \big> \nonumber \\
 &=& \f{\delta W}{\delta \ga(x\pr,x)},
 \label{eqn:w}
\end{eqnarray}  
which recovers Gilbert's result.
The kernel of the KS Hamiltonian may be written in the familiar form 
\begin{eqnarray}
 h(x,x\pr) &=& \delta(x-x\pr) \big(-\f{1}{2} \nabla_r^2 \big) 
 + v(x,x\pr)  \nonumber \\
 &&+ \delta(x-x\pr) v_H(x) + v_{xc}(x,x\pr),
 \label{eqn:1MFT:ham}
\end{eqnarray}
where $v(x,x\pr)$ is the external potential and $w(x,x\pr)$ 
has been divided into the Hartree $v_H(x)$ and 
exchange-correlation $v_{xc}(x,x\pr)$ potentials.  In 1MFT, 
the exchange-correlation potential is nonlocal.

The 1MFT-KS scheme optimizes the orbitals for a chosen set of 
occupation numbers, but it does not itself provide a rule for 
choosing the occupation numbers.
On this point, it is different from the DFT-KS scheme, where the occupation numbers 
are usually uniquely determined by the aufbau principle ($T=0$ Fermi 
statistics).\footnote{If the DFT-KS system is degenerate, the occupation numbers 
of the degenerate KS orbitals are not determined by the aufbau principle. In this
case, the KS system adopts an ensemble state, and the occupation numbers of 
the degenerate orbitals are chosen such that the KS system is self-consistent 
and reproduces the density of the interacting system.\cite{gross1990,ullrich2001}}
In 1MFT, the KS equations have a self-consistent solution for any chosen set of
occupation numbers $\{q_i\}$ that satisfy $0\leq q_i \leq 1$ and $\sum_i q_i = N$. 
The Lagrange multipliers $\epsilon_i$, which are seen to be the KS eigenvalues,
adopt values such that the minimum of $\mathcal{G}_v$ occurs for a 1-matrix 
$\ga_{min}$ whose occupation numbers are precisely the set $\{q_i\}$. 
Therefore, the unconstrained minimum of $\mathcal{G}_v$ 
coincides with the minimum of $E_v$ subject to the 
constraints $f_i=q_i$.   
To find the ground state occupation numbers, $\left\{q_i^{gs}\right\}$, one can 
search for the minimum of the function $G_v(\{q_i\})=\min_{\ga} \;\mathcal{G}_v[\ga]$, 
where the minimization of $\mathcal{G}_v[\ga]$  
can be performed by the KS scheme.  

What can be said about the KS eigenvalues $\epsilon_j$?  
As the minimum of $\mathcal{G}_v$ is a stationary point, we have
\begin{equation}
 0 = \left.\f{\partial \mathcal{G}_v}{\partial f_j}\right|_{\ga_{min}} 
 = \left.\f{\partial E_v}{\partial f_j}\right|_{\ga_{min}} - \epsilon_j(\{q_i\})
 \label{eqn:coll}
\end{equation}
for all $j$. $E_v$ is a variational functional in the ENR space.  
It attains its minimum at the ground state 1-matrix $\ga_{gs}$, where 
$\partial E_v/\partial f_j$ must vanish for all fractionally occupied 
($0<f_j<1$) orbitals, for otherwise the energy could be lowered. 
When $q_i=q_i^{gs}$ for all $i$, $\ga_{min}= \ga_{gs}$, and (\ref{eqn:coll}) implies 
$\epsilon_j(\{q_i^{gs}\})=0$ for all fractionally occupied orbitals.   
Thus, we find that the KS eigenvalues of all orbitals that are fractionally 
occupied in the ground state must collapse to a single level 
when the chosen set of occupation numbers approach their ground state 
values, i.e., as $q_i \rightarrow q_i^{gs}$.  
In Gilbert's derivation of the 1MFT-KS equations, all fractionally 
occupied KS orbitals were found to have the eigenvalue $\epsilon_i=\mu$, where
$\mu$ is the chemical potential.  As we have not introduced the
chemical potential in our derivation (we consider a system with a 
fixed number of electrons), the eigenvalues collapse to 
$0$ instead of $\mu$. The above arguments do not apply to orbitals 
with occupation numbers exactly $0$ or $1$ because these values lie 
on the boundary of the allowed interval $[0,1]$ specified by ENR 
condition (iii).  All that can be concluded from the fact that 
$E_v$ has a minimum in the ENR space is $\epsilon_j\geq 0$ for 
orbitals with $f_j=0$ and $\epsilon_j\leq 0$ for orbitals with $f_j=1$.  
States with occupation numbers exactly $0$ or $1$ have been called 
``pinned states.''\cite{helbig2007,lathiotakis2007}
Instances of such states in real systems have been reported,\cite{cioslowski2006} 
though their occurrence is generally considered to be exceptional.\cite{helbig2007,lathiotakis2007}

Due to the collapse of the eigenvalues at the ground state,
the KS Hamiltonian becomes the null operator
\begin{equation}
 \hat{h}[\gamma_{gs}] = \hat{0}
 \label{eqn:identity}
\end{equation}
in the subspace of fractionally occupied orbitals.  This is of 
course analogous to the familiar condition $df/dx=0$ for the 
extremum of a function $f(x)$.  Gilbert described\cite{gilbert1975} 
a similar result (with $\mu \hat{I}$ replacing $\hat{0}$) 
as ``paradoxical,'' a statement that has been repeated.\cite{valone1980,tal1985}
The problem with (\ref{eqn:identity}) is that while we expect the 
KS Hamiltonian to define the natural orbitals, any state is an 
eigenstate of the null operator.  However, the KS Hamiltonian is a 
functional of the 1-matrix, and when the occupation numbers are 
perturbed from their ground state values, the degeneracy is lifted 
and the KS Hamiltonian does define unique orbitals.  In the KS scheme 
outlined above, this corresponds to the optimization of the orbitals 
with occupation numbers fixed to values $q_i$, perturbed from the 
ground states values.  In the limit that the occupation numbers 
approach their ground state values, the optimal orbitals approach 
the ground state natural orbitals. The degenerate eigenvalues 
generally split linearly with respect to perturbations away from the 
ground state.  In particular,
\begin{eqnarray}
 \f{\partial \epsilon_i}{\partial q_j} &=& \int dy dy\pr 
 \left< \phi_i \right| \f{\delta \hat{h}}{\delta \ga(y,y\pr)} 
 \f{\partial \ga(y,y\pr)}{\partial q_j} \left| \phi_i\right> \nonumber \\
 &=& - \int dx dx\pr dy dy\pr \phi_i^*(x) \phi_j^*(y\pr) \chi^{-1}(x x\pr,y y\pr) \nonumber \\
 && \times \phi_i(x\pr) \phi_j(y). 
 \label{eqn:split}
\end{eqnarray}
Here, $\chi$ is the static response function defined as
\begin{equation}
 \chi(x,x\pr;y,y\pr) =  \f{\delta \gamma(x,x\pr)}{\delta v(y,y\pr)}.
 \label{eqn:response}
\end{equation}
The relation $\delta h/\delta \ga= -\chi^{-1}$ used in (\ref{eqn:split})
is derived in the following section. If $\chi$ has a null space, its 
inverse is defined only on a restricted space.  For example, 
(\ref{eqn:split}) does not apply to pinned states as there is a null 
vector associated with each pinned state (see below). 

Our derivation of the 1MFT-KS equations in fact assumes that the static
response function $\chi$ of the interacting system has no null vectors 
except for those associated with pinned states and a constant 
shift of the potential.  We now show that this guarantees $\mathcal{G}_v$ 
is stationary.  If the interacting system has any other null vectors we 
can no longer be certain that $\mathcal{G}_v$ is stationary and the 
derivation should be modified as described below.  We have remarked 
already (see Ref. 34) that $\mathcal{G}_v$ is stationary in the VR space, 
i.e., it satisfies the stationary condition $\delta \mathcal{G}_v=0$ with 
respect to an arbitrary variation of the 1-matrix in the VR space.  However, 
our derivation of the KS equations requires $\mathcal{G}_v$ to be stationary 
in the ENR space.  As the VR space is a subspace of the ENR space, this 
is a stronger condition.  The assumption that $\chi$ has no null vectors 
(apart from those associated with pinned states) is equivalent to assuming 
that any ENR variation (apart from variations of the pinned occupation 
numbers) is also a VR variation.  For if $\chi$ has no null vectors, then 
it is invertible and any ENR variation $\delta \hat{\ga}$ can be induced by 
the perturbation $\delta \hat{v} = \chi^{-1} \delta \hat{\ga}$.\footnote{We do
not demonstrate this statement for $\chi$ as an operator on an infinite 
dimensional space. However, it is valid in a finite dimensional basis 
(see Ref.~\onlinecite{kohn1983} for a proof in DFT).}
Hence, with the above assumption, $\mathcal{G}_v$ is guaranteed to be 
stationary with respect to any ENR variation.  The above arguments do not 
apply to variations of pinned occupation numbers because there are null 
vectors associated with such variations; nevertheless, 
$\mathcal{G}_v$ is stationary with respect to such variations as this 
is maintained by the Lagrange multipliers $\epsilon_i$.
It is now clear how to modify the derivation of the KS equations when 
$\chi$ has additional null vectors.  By introducing new Lagrange multipliers, 
the additional null vectors can be treated in analogy with the pinned states.  
For example, suppose $\chi$ has one additional null vector
$\hat{u}=\sum_{ij} u_{ij} \left| \phi_i \right> \left< \phi_j\right|$, 
where $\hat{u}$ is hermitian and $Tr(\hat{u}\hat{u})=1$.  
The new energy functional
$\mathcal{G}\pr_v[\ga] = \mathcal{G}_v[\ga] - \kappa (\ga_u-p_u)$
will be stationary with respect to an arbitrary variation in the ENR space.
Here, the Lagrange multiplier $\kappa$ enforces the constraint $\ga_u=p_u$, 
where $\ga_u=Tr(\hat{\ga}\hat{u})$ is the component of $\hat{\ga}$ 
corresponding to $\hat{u}$.  The stationary condition 
$\delta \mathcal{G}\pr_v=0$ leads to the set of equations 
$h_{ij}-\epsilon_j \delta_{ij} - \kappa u_{ij}=0$ in the basis of natural
orbitals.  The 1-matrix that satisfies these equations can be found by
solving self-consistently the eigenvalue equation 
$\hat{h}\left| \xi_i\right> = \omega_i \left|\xi_i\right>$ together 
with $\ga(x,x\pr)=\sum_{ijk} q_i S_{ji} S^*_{ki} \xi_j(x) \xi^*_k(x\pr)$, 
where $S$ is the unitary matrix that diagonalizes the matrix $u$ in the 
basis of natural orbitals, i.e., $S u S^{\dag}$ is diagonal.  
The energy of the self-consistent solution defines a 
function $G\pr_v(\{q_i\},p_u)$, whose minimum with respect to $\{q_i\}$ 
and $p_u$ is the ground state energy.  It may not be known in advance 
whether the response function of a given interacting system will have null 
vectors.  Therefore, it is helpful to understand how null vectors occur.

Null vectors of $\chi$ are connected with the so-called nonuniqueness 
problem\cite{vonbarth1972,eschrig2001,capelle2001} in various extensions
of DFT. A system with the ground state $\Psi_0$ is said to have a nonuniqueness 
problem if there is more than one external potential for which $\Psi_0$ is the
ground state.  The Schr\"odinger equation defines a unique map from the 
external potential to the ground state wavefunction (if it is nondegenerate), 
but when there is more than one external potential yielding the same ground 
state wavefunction, the map cannot be inverted.
In 1MFT the generality of the external potential (nonlocal in space and 
spin coordinates) allows greater scope for nonuniqueness than in the other
extensions of DFT.  Of course, every degree of nonuniqueness is a null vector 
of $\chi$ because if $\delta \hat{v}$ does not change the ground state 
wavefunction, it does not change the 1-matrix either, and hence it is a null 
vector.  In fact, every null vector of $\chi$ is caused by nonuniqueness;  
the existence of a null vector $\delta \hat{v}$ that induced a nonzero $\delta \Psi_0$
would contradict the one-to-one relationship $\ga \leftrightarrow \Psi_0$
proved by the extension of the HK theorem to 1MFT.\cite{gilbert1975}
It was mentioned above that there are null vectors associated with the 
pinned states.  Suppose $\phi_k$ is a
natural orbital with occupation number $f_k=0$ in the ground state.  
The perturbation of the external potential 
$\delta \hat{v} = \lambda \left| \phi_k \right>\left< \phi_k \right|$
does not change the ground state if the system has an energy gap between the
ground state and excited states and $\lambda$ is small enough because 
$\delta \hat{V} \Psi_0 = 0$, where 
$\delta \hat{V}= \int dx dx\pr \hat{\psi}^{\dag}(x) \delta v(x,x\pr) \hat{\psi}(x\pr)$
and $\hat{\psi}$ and $\hat{\psi}^{\dag}$ are field operators.
If $f_k=1$, $\delta \hat{V} \Psi_0 = \Psi_0$ and the ground state is 
again unchanged by the perturbation.  The ``vector''  
$\left| \phi_k \right>\left< \phi_k \right|$ 
is therefore a null vector of $\chi$ if $\phi_k$ is a pinned state.  
Another type of nonuniqueness, which has been called systematic 
nonuniqueness,\cite{capelle2001} is related to constants of the 
motion.  Suppose 
$\hat{A} = \int dx dx\pr \hat{\psi}^{\dag}(x) a(x,x\pr) \hat{\psi}(x\pr)$ 
is a constant of the motion.  The ground state, if it is nondegenerate,
is an eigenstate of $\hat{A}$ as constants of the motion commute with the 
Hamiltonian.  If the system has an energy gap between the ground state and 
the first excited state, then a perturbation $\delta \hat{V} =\lambda \hat{A}$  
will not change the ground state wavefunction if $\lambda$ is small 
enough.  Thus, $\hat{A}$ is a null vector of $\chi$.

As in DFT, an exact and explicit expression for the universal energy functional in 
1MFT is unknown in general.  In actual calculations it is usually necessary 
to use approximate functionals. 
Many of the approximate energy functionals that have been introduced are 
expressed in terms of the natural orbitals and occupation numbers.  Such 
functionals are valid 1-matrix energy functionals, but as the dependence 
on the 1-matrix is implicit rather than explicit, they have been called 
``implicit'' functionals.  Recently, the KS equations were derived for 
this case.\cite{pernal2005}  It was found that the contribution to the KS 
potential from electron-electron interactions can be evaluated by applying 
the following chain rule to (\ref{eqn:w}),
\begin{eqnarray}
 w(x,x\pr) &=& \f{\delta W}{\delta \ga(x\pr,x)} \nonumber \\
 &=& \sum_i \int dy \f{\delta W}{\delta \phi_i(y)} \f{\delta \phi_i(y)}{\delta \ga(x\pr,x)} \nonumber \\
 &+& \sum_i \int dy \f{\delta W}{\delta \phi_i^*(y)} \f{\delta \phi_i^*(y)}{\delta \ga(x\pr,x)} \nonumber\\
 &+& \sum_i \f{\partial W}{\partial f_i} \f{\delta f_i}{\delta \ga(x\pr,x)}.
 \label{eqn:chainB}
\end{eqnarray}

\subsection{\label{ssec:iter} Iteration of the KS equations}

In this section we show that the ``straightforward'' procedure for 
iterating the KS equations (\ref{eqn:1-matrix:diag}) and (\ref{eqn:1MFT-KS}) 
is intrinsically divergent.  

The KS equations are nonlinear because the KS Hamiltonian 
itself depends on the 1-matrix. In favorable cases such nonlinear 
equations can be solved by iteration.  Given a good initial guess 
for the 1-matrix, iteration may lead to the self-consistent 
solution corresponding to the ground state.
In order to iterate (\ref{eqn:1MFT-KS}), one needs an algorithm to 
define the 1-matrix of iteration step $n+1$ from the 1-matrix 
of step $n$, i.e., one needs to ``close'' the KS equations. In 
the previous section, we saw that in the 1MFT-KS scheme the 
occupation numbers $f_i$ are held fixed during the optimization 
of the natural orbitals.  The following is a ``straightforward'' 
algorithm that optimizes the natural orbitals: i) the KS 
Hamiltonian for step $n+1$ is defined by
\begin{equation}
 \hat{h}^{(n+1)} = \hat{h}[\gamma^{(n)}],
 \label{eqn:ham:step}
\end{equation}
where $\hat{h}$ is given by (\ref{eqn:1MFT-ham}) 
and $\gamma^{(n)}$ is the 1-matrix of iteration step $n$;
ii) the eigenstates of $\hat{h}^{(n+1)}$ are taken as 
the natural orbitals of step $n+1$;  
iii) the 1-matrix of step $n+1$ is constructed from the 
natural orbitals of step $n+1$ by the expression
\begin{equation}
 \gamma^{(n+1)}(x,x\pr) = \sum_i f_i \phi_i^{(n+1)}(x) \phi_i^{*(n+1)}(x\pr).
\end{equation}
Let $u_i$ be the eigenstates of the KS Hamiltonian $\hat{h}^{(n+1)}$.
In operation (ii), the natural orbitals $\phi_i^{(n+1)}$ are
chosen from among the $u_i$ such that $\phi_i^{(n+1)}$ has maximum 
overlap with $\phi_i^{(n)}$, i.e., $Tr((\hat{\ga}^{(n+1)}-\hat{\ga}^{(n)})^2)$ 
is the minimum possible.
If this procedure converges to the stationary 1-matrix 
$\gamma_{min}$ giving the lowest energy possible for the fixed set of 
occupation numbers, then it defines the function $G_v(\{f_i\})$, introduced
in the previous section, for which the ground state energy is the absolute minimum.  
Unfortunately, for any set of occupation numbers $\{f_i\}$ sufficiently 
close to the ground state occupation numbers, this procedure does not 
converge to $\gamma_{min}$.  In other words, the ``straightforward'' 
algorithm defines an iteration map for which the ground state $\gamma_{gs}$
is an unstable fixed point.  This is a consequence 
of the degeneracy of the KS spectrum at the ground state.

The divergence of the iteration map is revealed by a linear 
analysis of the fixed point.  
Suppose the occupation numbers are fixed to values perturbed 
from their ground state values by $\delta f_i$.  Let us consider 
an iteration step $n$ and ask whether the next iteration takes us 
closer to the stationary point $\gamma_{min}$ that gives the minimum 
energy for the fixed occupation numbers.  The linearization of the 
iteration map at the stationary point gives
\begin{eqnarray}
 \delta \hat{\gamma}^{(n+1)} 
 &=& \hat{\gamma}^{(n+1)} - \hat{\gamma}_{min} \nonumber \\
 &\approx& \hat{\chi}_s[\gamma_{min}] \big( \hat{v}_s^{(n+1)} - \hat{v}_s^{min} \big) \nonumber \\
 &=& \hat{\chi}_s[\gamma_{min}] \big( \hat{h}^{(n+1)} - \hat{h}^{min} \big) \nonumber \\
 &\approx& - \hat{\chi}_s[\gamma_{min}] \hat{\chi}^{-1} \delta \hat{\gamma}^{(n)},
 \label{eqn:map:lin}
\end{eqnarray}
where $\hat{v}_s=\hat{v}+\hat{v}_H+\hat{v}_{xc}$ is the KS 
potential. The response function $\chi$ was defined in 
(\ref{eqn:response}). The KS response function is
\begin{eqnarray}
 \chi_s(x,x\pr;y,y\pr) 
 &=& \f{\delta \gamma(x,x\pr)}{\delta v_s(y,y\pr)} \nonumber \\
 &=& \sum_i \sum_{j\neq i} \f{f_i -f_j}{\epsilon_i-\epsilon_j} \nonumber \\
   &\times& \phi_j(x) \phi_i^*(x\pr) \phi_j^*(y) \phi_i(y\pr).
 \label{eqn:KS:response}
\end{eqnarray}
In the last line of (\ref{eqn:map:lin}), we have used
\begin{eqnarray}
 \hat{h}^{(n+1)} - \hat{h}^{min} &\approx& -\hat{\chi}^{-1} \delta \hat{\gamma}^{(n)},
 \label{eqn:ham:variation}
\end{eqnarray}
which can be established by the following arguments.  First consider 
\begin{eqnarray}
 \hat{h}^{(n+1)} - \hat{h}^{min} 
 &\approx& \left. \f{\delta \hat{h}[\ga]}{\delta \gamma}\right|_{\gamma_{min}} \delta \hat{\gamma}^{(n)},
 \label{eqn:ham:variation2}
\end{eqnarray}
which follows from (\ref{eqn:ham:step}). The KS Hamiltonian is an
implicit functional of the 1-matrix, and (\ref{eqn:ham:variation2}) 
defines the first order change of the KS Hamiltonian with respect to 
a perturbation of that 1-matrix.   Since the occupation numbers are 
close to their ground state values, we may make the replacement
\begin{equation}
 \left. \f{\delta \hat{h}}{\delta \gamma}\right|_{\gamma_{min}} 
 \rightarrow \left. \f{\delta \hat{h}}{\delta \gamma}\right|_{\gamma_{gs}},
 \label{eqn:deriv}
\end{equation}
which is valid to $\mathcal{O}(\max(|\delta f_i|))$. 
Thus, to establish (\ref{eqn:ham:variation}) we need to show 
$\delta \hat{h}/\delta \gamma = - \hat{\chi}^{-1}$ at the 
ground state 1-matrix $\ga_{gs}$.  
The KS Hamiltonian $\hat{h}$ is associated with the original 
many-body Hamiltonian $\hat{H}$, which has external potential $v(x,x\pr)$.
According to (\ref{eqn:identity})\footnote{For convenience we assume here
that the system has no pinned states.}, $\hat{h}[\gamma_{gs}]=\hat{0}$.  
Consider now a Hamiltonian $\hat{H}\pr$ with a slightly 
different external potential 
$v\pr(x,x\pr)=v(x,x\pr)+\delta v(x,x\pr)$ such that 
its ground state 1-matrix is $\gamma\pr_{gs}=\ga_{gs}+\delta \ga$.
The associated KS Hamiltonian is 
$\hat{h}\pr = \hat{h}+\hat{v}\pr-\hat{v}$.  At
the new ground state, $\hat{h}\pr[\gamma_{gs}\pr] = \hat{0}$.
This allows us to relate $\delta \hat{h}$ to $\delta \hat{v}$ as
\begin{eqnarray}
 \delta \hat{h} 
 &=& \hat{h}[\gamma_{gs}\pr] - \hat{h}[\gamma_{gs}] \nonumber \\
 &=& \hat{h}[\gamma_{gs}\pr] + \delta \hat{v} - \delta \hat{v}  \nonumber \\
 &=&  - \delta \hat{v}.
\end{eqnarray}
Finally, using $\delta \hat{v} = \hat{\chi}^{-1} \delta \hat{\gamma}$ 
we obtain $\delta \hat{h}/\delta \gamma = - \hat{\chi}^{-1}$, which verifies
(\ref{eqn:ham:variation}).

Returning to the question of convergence, we see that
(\ref{eqn:map:lin}) implies that the next iteration 
takes us farther from the stationary point $\gamma_{min}$.  
The reason is that $\left|\det \hat{\chi}_s \hat{\chi}^{-1}\right|>1$ if 
$\gamma$ is sufficiently close to the ground state. According 
to (\ref{eqn:split}) the KS response diverges as 
$\gamma \rightarrow \gamma_{gs}$ because 
$\epsilon_i-\epsilon_j \sim \mathcal{O}(\max(|\delta f_i|))$.
For a fixed set of occupation numbers sufficiently close to 
their ground state values, the moduli of all eigenvalues of the
operator $\hat{\chi}_s \hat{\chi}^{-1}$ become greater than 1 
(the null space of $\chi$ is assumed to be excluded).
Therefore, a perturbation $\delta \gamma$ from the ground state 
is amplified by iteration, and the ground state is an unstable
fixed point of the iteration map.  A fixed point is 
stable if and only if all eigenvalues of the linearized
iteration map have modulus less than 1.

\subsection{\label{ssec:shifting} Level shifting method}

In the previous section, it was shown that the ``straightforward'' iteration
of the KS equations is intrinsically divergent.  To obtain a practical 
KS scheme the iteration map must be modified.  In this section, we 
consider the level shifting method\cite{saunders1973} and by linearizing 
the modified iteration map we obtain a criterion for convergence.

Intrinsic divergent behavior can be encountered also in the Hartree-Fock 
approximation, and various modifications of the iteration procedure have 
been introduced, for example, Hartree damping (also called configuration
mixing) and level shifting.  The level shifting method is particularly 
attractive in 1MFT because it can prevent the collapse of eigenvalues 
that is the origin of divergent behavior.  Indeed, it has been shown 
that the level shifting method is capable of giving a convergent 
KS scheme in 1MFT.\cite{pernal2005}

In the ``straightforward'' iteration procedure, the change of the orbitals, 
to first order, from iteration step $n$ to iteration step $n+1$ is 
\begin{eqnarray}
 \delta \phi_i(x) &=& \phi_i^{(n+1)}(x) - \phi_i^{(n)}(x) \nonumber\\
 &=& \sum_{j\neq i} 
\frac{\left<\phi_j \right| \hat{h}^{(n+1)} - \hat{h}^{(n)} \left| \phi_i \right>}{\epsilon_i-\epsilon_j} \phi_j(x),
\label{eqn:first}
\end{eqnarray}
where $\hat{h}^{(n)}$ is the KS Hamiltonian for iteration step $n$ 
and in the second line the orbitals and eigenvalues are from iteration 
step $n$.  In the level shifting method, 
the first order change in the orbitals given by (\ref{eqn:first}) is altered by 
applying the shifts $\epsilon_i \rightarrow \epsilon_i + \zeta_i$ to the eigenvalues 
in the denominator.  To first order, this modification is equivalent to adding 
the term $\hat{\Delta} = \sum_i \zeta_i \big| \phi_i^{(n)} \big> \big< \phi_i^{(n)} \big|$ 
to the KS Hamiltonian for step $n+1$. Let $\hat{h}_{\zeta} = \hat{h} + \hat{\Delta}$ 
define the level shifted Hamiltonian.  Repeating the linear analysis of the 
previous section for the iteration map for this level shifted Hamiltonian, we find
\begin{eqnarray}
 \hat{\delta \ga}^{(n+1)} 
 &\approx& \hat{\chi}_s[\gamma_{min}] \big( \hat{v}_s^{(n+1)} - \hat{v}_s^{min} \big) \nonumber \\
 &=& \hat{\chi}_s[\gamma_{min}] \big( \hat{h}_{\zeta}^{(n+1)} - \hat{h}_{\zeta}^{min} \big) \nonumber \\
 &\approx& \big( -\hat{\chi}_s[\gamma_{min}] \hat{\chi}^{-1} + \hat{\Omega} \big) \hat{\delta \ga}^{(n)},
 \label{eqn:modmap:lin}
\end{eqnarray}
where we have defined the operator $\hat{\Omega}$ with the kernel
\begin{eqnarray}
 \Omega(x x\pr, y y\pr) &=& \int dz dz\pr \chi_s(x x\pr, z z\pr) 
 \f{\delta \Delta(z z\pr)}{\delta \ga(y y\pr)}  \nonumber \\
 &=& \sum_i \sum_{j\neq i} \phi_j(x) \phi_i^*(x\pr) \phi_j^*(y) \phi_i(y\pr). 
\end{eqnarray}
From the last line of (\ref{eqn:modmap:lin}), we obtain a criterion for 
the convergence of the iteration map.  All of the eigenvalues of the operator 
\begin{equation}
 \hat{\mathcal{A}} = -\hat{\chi}_s[\gamma_{min}] \hat{\chi}^{-1} + \hat{\Omega}
 \label{eqn:criterion}
\end{equation}
must have modulus less than 1.  The dependence on the level shift 
parameters $\zeta_i$ enters only through the shifted eigenvalues in the 
denominator of $\chi_s$.  The level shifting method is effective 
because it prevents
the divergence of $\chi_s$ at the ground state and there is 
a cancellation between the two terms in (\ref{eqn:criterion}).  
Unfortunately, the convergence criterion depends on $\chi$, which 
is unknown at the outset of a 1MFT calculation.  In Sec.~\ref{sec:Hubbard}, 
the level shifting method is applied in an
explicit example and the above criterion is verified.

\subsection{\label{ssec:props} Properties of the KS system}

The distinguishing feature of the KS system in 1MFT is the
degeneracy of the eigenvalue spectrum.  This has 
surprising consequences.  It was shown in section \ref{ssec:defn} 
that the KS eigenvalue spectrum splits linearly as we move away from 
the ground state 1-matrix.  Therefore, the total KS energy changes  
linearly with respect to the displacement, i.e., 
$E_s[\gamma]-E_s[\gamma_{gs}] \propto \delta \gamma$, where 
$E_s[\gamma]=tr(\hat{h}[\gamma] \gamma)$ (for a specific example 
see Fig. \ref{fig:curve} in Sec.~\ref{sssec:1MFT:iter}).  
This is surprising because such linear changes do not occur for the 
energy functional $E_v$ (in the VR space).
The immediate implication is that $E_s[\ga]$ is not stationary at the 
ground state.  
While this causes no difficultly in principle --- we need only 
the functional $E_v$ to be stationary --- it is intimately
connected with the divergence of the iteration map.
Precisely at the ground state $E_s[\gamma_{gs}] = \sum_i\pr \epsilon_i$,
where the prime indicates that only the pinned states with $f_i=1$ 
contribute to the sum. Away from the ground state the KS eigenvalue 
spectrum splits, and $E_s[\gamma]$ is a multivalued functional due
to the choice implied in occupying the new KS levels.  This 
is the same choice encountered in the iteration of the KS
equations (see Sec.~\ref{ssec:iter}), where the natural orbitals 
$\phi_i^{(n+1)}$ are selected from among the eigenstates of the 
KS Hamiltonian. Near the self-consistent solution, there will be 
one such choice for which the resulting $\ga^{(n+1)}$ is very 
close to $\ga^{(n)}$.

It was shown in the preceding section that the static 
response function of the KS system diverges  
at the ground state.  Thus, even an infinitesimal perturbation 
$\delta \hat{v}_s$ may induce a finite change of 
$\gamma$.  At the ground state, all of the natural orbitals,
except those which have an occupation number that is 
degenerate, are uniquely defined.  The natural orbitals 
which belong to a degenerate occupation number are only defined modulo 
unitary rotation in the degenerate subspace.  When a 
perturbation is introduced, the natural orbitals change 
discontinuously to the eigenstates of the perturbed KS 
Hamiltonian $\hat{h} = \delta\hat{v}$.  These eigenstates may
be \emph{any} functions in the degenerate Hilbert space because
$\delta \hat{v}$ is arbitrary. 


\section{\label{sec:Hubbard} Two-site Hubbard model}

The 1MFT-KS system has some unusual features, such as
the collapse of the KS eigenvalues at the ground state, 
so it is desirable to derive explicitly the KS equations 
for a simple model.  The Hubbard model on two sites provides 
a convenient example because it is exactly solvable and 
especially easy to interpret.   
Also, analytic expressions for the 1-matrix energy 
functional and KS Hamiltonian can be obtained.
In the following sections, for the purpose of 
comparison, we find the ground state of the two-site Hubbard 
model by three methods --- direct solution of the Schr\"odinger 
equation, 1MFT and DFT. 

\subsection{\label{ssec:direct} Direct solution}

The Hamiltonian of the two-site Hubbard model is
$\hat{H} = \hat{T} + \hat{U} + \hat{V}$ with
\begin{eqnarray}
 \hat{T} &=& - \sum_{\s} \LB t_{12} c_{1\s}^{\dag} c_{2\s} + t_{21} c_{2\s}^{\dag} c_{1\s} \RB \nonumber \\
 \hat{U} &=& U \LB \hat{n}_{1\u} \hat{n}_{1\d} + \hat{n}_{2\u} \hat{n}_{2\d} \RB \nonumber \\
 \hat{V} &=& V \f{1}{2} \LB \hat{n}_1 - \hat{n}_2 \RB, 
 \label{eqn:Hamiltonian2}
\end{eqnarray}
where $t_{12}=t_{21}=t$, $c_{i\s}^{\dag}$ and $c_{i\s}$ are the 
creation and annihilation operators of an electron at site 
$i$ with spin $\s$, and $\hat{n}_i=\sum_{\s} c_{i\s}^{\dag} c_{i\s}$.
We consider only the sector of states with $N=2$ and $S_z=0$, 
i.e., a spin unpolarized system.  In this sector, the 
eigenstates of $\hat{T}+\hat{U}$ are
\begin{equation}
 \begin{array}{ll}
 \Phi_0 = \f{1}{\sqrt{2}} \LB \begin{array}{r} y \\ x \\ x \\ y \end{array} \RB,  &
 \Phi_1 = \f{1}{\sqrt{2}} \LB \begin{array}{r} 0 \\ 1\\ -1\\ 0 \end{array} \RB,  \\
 \Phi_2 = \f{1}{\sqrt{2}} \LB \begin{array}{r} 1 \\ 0 \\ 0 \\ -1 \end{array} \RB, &
 \Phi_3 = \f{1}{\sqrt{2}} \LB \begin{array}{r} x \\ -y \\ -y \\ x \end{array} \RB, 
\end{array}
\label{eqn:Phi}
\end{equation}
in the site basis $\big\{c_{1\u}^{\dag} c_{1\d}^{\dag} \left| 0\right>$, 
$c_{1\u}^{\dag} c_{2\d}^{\dag} \left| 0\right>$, 
$c_{2\u}^{\dag} c_{1\d}^{\dag} \left| 0\right>$,
$c_{2\u}^{\dag} c_{2\d}^{\dag} \left| 0\right>\big\}$.
The following variables have been introduced 
$x=\cos(\pi/4-\alpha_0/2)$, $y=\sin(\pi/4-\alpha_0/2)$, and 
$\tan \alpha_0 = U/4t$ with $0\leq \alpha_0 \leq \pi/2$.  The 
eigenvalues of $\hat{T}+\hat{U}$ for the states $\Phi_i$ are 
\begin{eqnarray*}
 &\lambda_0 = -B y^2, \quad \lambda_1 = 0 \\
 &\lambda_2 = B(x^2-y^2), \quad \lambda_3 = B x^2,
\end{eqnarray*}
where $B=\sqrt{U^2+T^2}$ and $T=4t$.
$\Phi_0$, $\Phi_2$, and $\Phi_3$ are singlet states ($S=0$) and $\Phi_1$ is
a triplet state with $S=1$ and $S_z=0$.  We will omit $\Phi_1$ from consideration
as it is not coupled to the other states by the spin-independent external
potential chosen in (\ref{eqn:Hamiltonian2}).  The Hamiltonian may be written 
as $\hat{H} = \lambda_0 \hat{I} + B \hat{K}$, where 
\begin{equation}
 K = \LB 
\begin{array}{ccc}
0      & \nu y   &   0   \\
\nu y  &  x^2    & \nu x \\
0      & \nu x   &   1
\end{array} \RB
\end{equation}
in the basis $(\Phi_0,\Phi_2,\Phi_3)$.  We have defined the 
dimensionless variable $\nu = V/B$.  The secular equation
$\big| \hat{K}-\kappa_i \hat{I}\big|=0$ is
\begin{equation}
 \kappa_i^3-(x^2+1) \kappa_i^2+(x^2-\nu^2) \kappa_i + \nu^2 y^2 = 0.
 \label{eqn:secular}
\end{equation}
The normalized eigenvectors of $\hat{H}$ are
\begin{equation}
 \Psi_i= \f{1}{\eta_i} \LB \begin{array}{c} \nu^2 x y \\ \nu x \kappa_i \\ \beta_i  \end{array}\RB
 \label{eqn:Psi}
\end{equation}
and have energy $E_i = \lambda_0 + B\kappa_i$ for $i=0,2,3$, where 
$\kappa_i$ is a root of the secular equation (\ref{eqn:secular}).
We have also defined 
\begin{equation}
 \beta_i = \kappa_i (\kappa_i-x^2)-\nu^2 y^2
\end{equation}
and
\begin{eqnarray}
 \eta_i &=& \left[\kappa_i^2 (\nu^2(3x^2-1)+y^2) + \kappa_i (x^2 y^2 (2\nu^2-1) )\right. \nonumber \\
 &&\left.+ \nu^2 y^2 (\nu^2-y^2) \right]^{1/2}.
\end{eqnarray}

\begin{figure}[ht!]
\centering
\includegraphics[width=0.9\columnwidth]{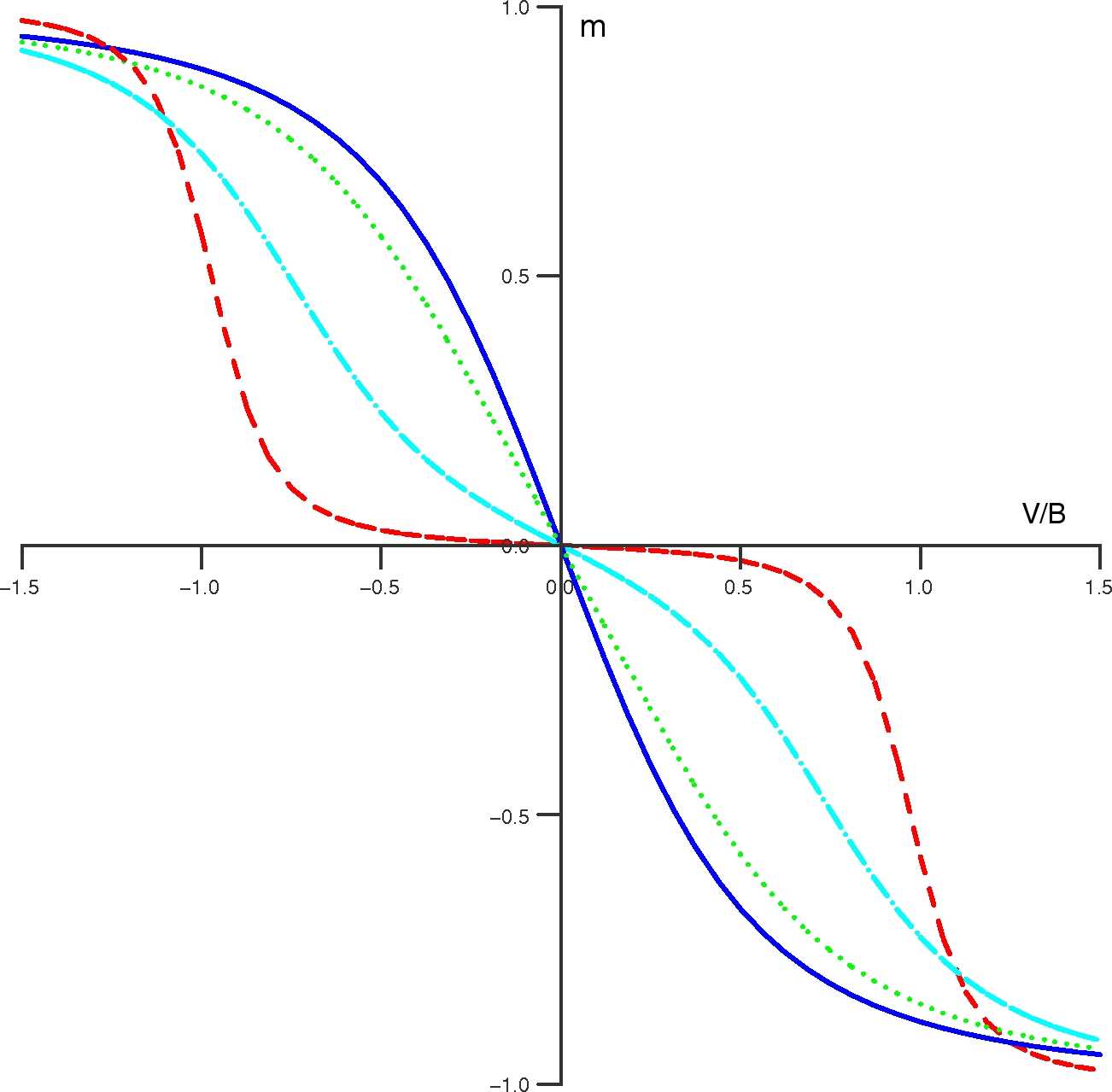}
\caption{\label{fig:m} [color online] The density variable $m=(n_1-n_2)/2$, 
is shown with respect to the dimensionless external potential 
$\nu=V/B$ ($B=\sqrt{T^2+U^2}$). Curves for $U/T=(1/16,1/4,1,4)$ 
are shown as (solid [blue], dotted [green], dash-dotted 
[light blue], dashed [red]) curves, respectively.}
\end{figure}

The two dimensionless energy scales of the system are 
the interaction strength $U/T$ and the bias $V/T$.  The 
behavior of the system with respect to these energy scales 
is illustrated in Fig.~\ref{fig:m}. The quantity 
$m=(n_1-n_2)/2$, where $n_i$ is the average ground state 
occupancy of site $i$, is plotted with respect to the 
external potential $V$ for various values of the 
interaction strength $U$.
For the ground state,  
\begin{eqnarray}
 m &=& \left< \Psi_0 \left| \hat{m} \right| \Psi_0 \right> \nonumber \\
   &=& \f{2\nu \kappa_0^2 x^2}{\eta_0^2} \LB \kappa_0 - x^2 \RB,
 \label{eqn:mV}
\end{eqnarray}
where $\hat{m} = (\hat{n}_1-\hat{n}_2)/2$.
A weakly interacting system (e.g., the solid [blue] curve 
in Fig.~\ref{fig:m}) responds strongly to the
external potential.  In contrast, a strongly
interacting system (e.g., the dashed [red] curve) responds 
weakly up to a threshold $V/B \sim 1$ (for 
a strongly interacting system $B \approx U$.)
This behavior has a simple interpretation:
in order for the external bias to induce 
charge transfer, it must overcome the on-site 
Hubbard interaction.
In the limit $U\rightarrow \infty$, the curve develops 
step-like behavior near $V/B \sim \pm 1$.

\subsection{\label{ssec:1MFT} Solution by 1MFT}

In the first part of this section, we derive the energy
functional and KS Hamiltonian.  In the second, we 
demonstrate the divergence of the iteration of the
KS equations. In the third, we use the level shifting 
method \cite{saunders1973} to obtain a convergent KS
scheme.

\subsubsection{\label{sssec:1MFT:E} Energy functional and KS Hamiltonian}

For lattice models such as the Hubbard model, the 
1-matrix is defined as
\begin{equation}
 \gamma(i\sigma,j\tau) = \big< \Psi \big| c_{i\sigma}^{\dag} c_{j\tau} \big| \Psi \big>.
 \label{eqn:1-matrix:lattice}
\end{equation}
One may ask whether the HK theorem (or Gilbert's extension in 1MFT) 
applies when the density (or 1-matrix) is defined over a discrete 
set of points, i.e., when the continuous density function $n(r)$ is 
replaced by the site occupation numbers $n_i$.  This has been
investigated, \cite{gunnarsson1986,xianlong2006} and it 
was found that the HK theorem remains valid.
We consider here only spin unpolarized states ($S_z=0$).  
Accordingly, we define the \emph{spatial} 1-matrix
\begin{equation}
 \gamma(ij) = \sum_{\sigma} \gamma(i\sigma,j\sigma). 
 \label{eqn:1-matrix:spatial}
\end{equation}
The 1-matrix may be expressed as, cf. (\ref{eqn:1-matrix:diag}),
\begin{equation}
 \gamma(ij) = \sum_{\alpha} f_{\alpha} \phi_{\alpha}(i) \phi_{\alpha}^*(j),
 \label{eqn:1-matrix:discrete}
\end{equation}
where $\phi_{\alpha}$ are the spatial natural orbitals.
As our system is spin unpolarized, 
the spin up and spin down spin-orbitals have the same
spatial factors.  Therefore, in (\ref{eqn:1-matrix:discrete})
each spatial orbital $\phi_{\alpha}$ may be occupied
twice (once by a spin up electron and once by a spin down 
electron), i.e., $0\leq f_{\alpha}\leq 2$. It is convenient
to parametrize the natural orbitals as
\begin{equation}
 \phi_a = \LB \begin{array}{rr} \cos(\theta/2) \\ \sin(\theta/2) \end{array} \RB  \quad \mathrm{and} \quad
 \phi_b = \LB \begin{array}{rr} \sin(\theta/2) \\ -\cos(\theta/2) \end{array} \RB. 
 \label{eqn:orbitals}
\end{equation}
In terms of this parametrization, the 1-matrix in the site
basis is
\begin{eqnarray}
 \gamma &=&  I + A \LB \cos \theta \s_z + \sin \theta \s_x \RB \nonumber \\
 &=&  I + \vec{\gamma} \cdot \vec{\s}; \quad \vec{\ga} = (\ga_x,\ga_z)
 \label{eqn:1-matrix:param}
\end{eqnarray}
where $\s_i$ are the Pauli matrices and $A= (f_a-f_b)/2 = \cos \alpha$.

For the two-site Hubbard model in the sector of singlet states
with $N=2$ and $S_z=0$, (\ref{eqn:1-matrix:lattice}) may be 
inverted to express $\Psi_0 = \Psi_0[\gamma]$.  Explicitly,
we find $\Psi_0 = \cos(\alpha/2) \Phi_{aa} - \sin(\alpha/2) \Phi_{bb}$,
where $\Phi_{ii}$ is the Slater determinant composed of the natural 
spin orbitals $\phi_{i\u}$ and $\phi_{i\d}$ ($i=a,b$).
The terms of the energy functional 
$E[\gamma] = T[\gamma] + U[\gamma] + V[\gamma]$ are  
found to be
\begin{eqnarray}
 T[\gamma] &=& \left< \Psi_0 \left| \hat{T} \right| \Psi_0 \right> = -2t A \sin \theta \nonumber \\ 
 U[\gamma] &=& \left< \Psi_0 \left| \hat{U} \right| \Psi_0 \right> = U - \f{U}{2} \LB 1+\sqrt{1-A^2} \RB \sin^2 \theta \nonumber \\
 V[\gamma] &=& \left< \Psi_0 \left| \hat{V} \right| \Psi_0 \right> = V A \cos \theta.
 \label{eqn:energy:fl}
\end{eqnarray}
The electron-electron interaction energy functional $U[\gamma]$
agrees with the general exact result for 2-electron closed
shell systems \cite{shull1959,kutzelnigg1963}.  We may 
partition $U[\gamma]$ into the 
Hartree energy 
\begin{eqnarray}
 E_H[\gamma] &=& \f{1}{2} \sum_{ij} n_i n_j U \delta_{ij} \nonumber \\
 &=& U \LB 1 + A^2 \cos^2 \theta \RB
\end{eqnarray}
and the exchange-correlation energy
\begin{eqnarray}
 E_{xc}[\gamma] &=& U[\gamma]- E_H[\gamma] \nonumber\\
 &=& -U \LB \f{1}{2}+\sqrt{1-A^2} \RB \nonumber \\
 &-& \f{U}{2} \LB 1 + A^2 + \sqrt{1-A^2} \RB \cos^2 \theta.
\end{eqnarray}

In Sec.~\ref{ssec:defn} the KS Hamiltonian was derived from the
stationary principle for the energy.  For the present model the KS 
Hamiltonian is a real $2\times 2$ matrix.  In the site basis its 
elements are
$h(ij) = \big<0 \big|c_i \hat{h} c_j^{\dag} \big| 0\big>$.
This matrix may be expressed as $h = \vec{h}\cdot \vec{\s}$ with
\begin{eqnarray}
 h_x &=& -\f{B}{4} \LB \cos \alpha_0 - \f{\sin \alpha_0}{\sin \alpha} \cos \alpha \sin \theta \RB \nonumber \\
      && -\f{B}{4} \f{\sin \alpha_0 (1+\sin \alpha)^2}{\sin \alpha \cos \alpha} \sin \theta \cos^2 \theta \nonumber\\
 h_y &=& 0  \nonumber\\
 h_z &=& \f{B}{4} \f{\sin \alpha_0 (1+\sin \alpha)^2}{\sin \alpha \cos \alpha} \sin^2 \theta \cos \theta
        +\f{V}{2}.
\label{eqn:h:pauli}
\end{eqnarray}
In these expressions the variable $\alpha$ represents the dependence 
on the occupation numbers through the definition 
$\alpha = \cos^{-1} A = \cos^{-1}((f_a-f_b)/2)$, $\alpha_0 = \tan^{-1}(U/4t)$
is the ground state value of $\alpha$ when $V=0$, and $\theta$ represents 
the dependence on the natural orbitals, c.f. (\ref{eqn:orbitals}).  
Let us verify (\ref{eqn:identity}) for the uniform case $V=0$, for
which the ground state 1-matrix has $\theta=\pi/2$ and 
$\alpha = \alpha_0$.  At these values $h_x=h_y=h_z=0$, which 
verifies the eigenvalue collapse in this case.  

\subsubsection{\label{sssec:1MFT:iter} Iteration of the KS equations}

\begin{figure}[ht!]
\includegraphics[width=0.95\columnwidth]{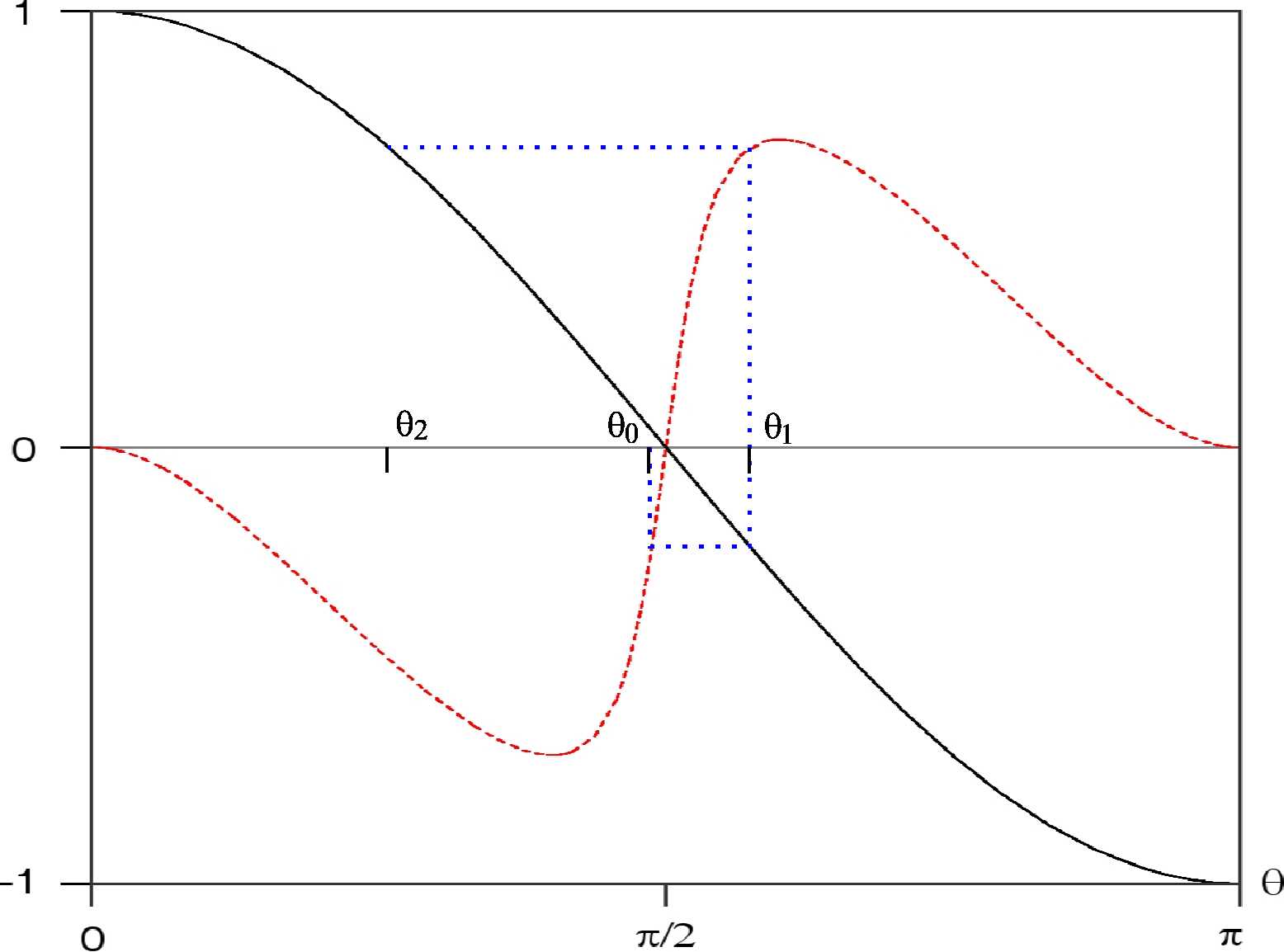}
\caption{\label{fig:map:1} [color online] The iteration map (\ref{eqn:map}) 
is shown for $t=1$, $U=1$, $V=0$, and $A=A_{gs}-0.02$. 
The solid [black] curve is the left hand side of (\ref{eqn:map}).  
The dashed [red] curve is the right hand side. The dotted [blue] curve 
demonstrates the first two iterations.  The iteration map 
does not converge to the ground state fixed point  
$\theta=\pi/2$.}
\end{figure}

We demonstrate here the iteration of the KS equations
following the straightforward algorithm described in Sec.~\ref{ssec:iter}.
During the optimization of the orbitals the occupation numbers
(i.e. $\alpha$) are held fixed.  Let us look more closely
at each operation in the algorithm.  In 
operation (i), the KS Hamiltonian for step $n+1$ is found
by evaluating (\ref{eqn:h:pauli}) at the 1-matrix 
$\gamma^{(n)}$, i.e., at $\theta = \theta_n$.  In operation
(ii), we find the eigenstates $u_i$ of $\hat{h}^{(n+1)}$, which we 
parametrize in the form (\ref{eqn:orbitals}) 
with $\theta=\theta_{n+1}$.  These eigenstates are taken as 
the natural orbitals $\phi_i^{(n+1)}$ for step $n+1$.
This implies setting each of the $\phi_i^{(n+1)}$
equal to one of the $u_i$.  In the present case, the natural
orbitals are chosen such that $\theta_{n+1}$ is as close as 
possible to $\theta_n$.
In operation (iii), $\gamma^{(n+1)}$ is constructed from the
$\phi_i^{(n+1)}$ by (\ref{eqn:1-matrix:discrete}).  We 
may now condense these three operations into a discrete 
iteration map on $\theta$, i.e., a map
$\theta_n \rightarrow \theta_{n+1}$.  It is defined by
\begin{equation}
 \cos \theta_{n+1} = \mathrm{sgn}(A-A_{gs}) \left. \f{h_z}{\sqrt{h_x^2+h_z^2}}\right|_{\theta=\theta_{n}}
 \label{eqn:map}
\end{equation}
for $0 < \theta_n < \pi$ and $A>0$ ($0<\alpha<\pi/2$). In 
(\ref{eqn:map}), $A_{gs}$ is the ground state value of $A$.
An example of the iteration map for $t=1$, $U=1$, $V=0$, 
and $A=A_{gs}-0.02$ is shown in Fig. \ref{fig:map:1}.  The 
solid [black] and dashed [red] curves are the left and right 
hand sides of (\ref{eqn:map}). The intersections of the two 
curves are fixed points of the iteration map. The ground state 
corresponds to the fixed point at $\theta=\pi/2$.  
The iteration map may be represented graphically by 
alternately drawing vertical lines from the solid curve to 
the dashed curve and horizontal lines from the dashed curve to the 
solid curve.  The dotted [blue] curve shows an example of the first
two iterations beginning from an initial guess $\theta_0$.
The next iterations $\theta_1$ and $\theta_2$ move farther away from the 
ground state, and the map does not converge to the ground
state fixed point $\theta=\pi/2$.

The iteration map is nonlinear and may exhibit quite complex 
behavior.  The linearization of the map at a fixed point
tells us whether the fixed point is stable or unstable.
As an example, let us consider the uniform case $V=0$, for
which the ground state fixed point is $\theta = \pi/2$.  
Linearization of (\ref{eqn:map}) in terms of the
variable $m = (n_1-n_2)/2 = A \cos \theta$ gives 
\begin{eqnarray}
 m_{n+1} &\approx& \mathrm{sgn}(A-A_{gs})\f{h_z}{\left|h_x\right|} \nonumber \\
 &\approx& -\xi m_n,
 \label{eqn:m:lin}
\end{eqnarray}
where 
\begin{equation}
 \xi = \f{(1+\sin \alpha)^2}{(\cot \alpha_0 - \cot \alpha) \sin \alpha \cos \alpha}.
\end{equation}
Suppose the occupation numbers are close to their ground state 
values, i.e., $A = A_{gs} + \delta A$ 
where $\delta A$ is a small displacement.  The leading 
approximation for $\xi$ gives
\begin{equation}
 \xi \approx -\f{U^2 (U+B)^2}{T B^3} \f{1}{\delta A}.
 \label{eqn:xi:val}
\end{equation}
For any nonzero values of $t$ and $U$, there is a threshold 
$d>0$ such that for 
$\left| \delta A \right| < d$, 
$\left|\xi \right| > 1$.  Therefore, the ground state 
is an unstable fixed point. In Sec.~\ref{ssec:iter}, 
the divergence of the iteration map was connected to 
the divergence of the static KS response function.  
Let us verify (\ref{eqn:map:lin}) explicitly for the 
present case.  
As seen in (\ref{eqn:m:lin}), the linearized iteration 
map affects only the diagonal elements of the 1-matrix, 
i.e., the density, which is described by the variable $m$.  
Therefore, the relevant response functions are the 
density-density response for the KS system
\begin{eqnarray}
 \chi_s &=& \sum_i \sum_{j\neq i} \frac{f_i - f_j}{\epsilon_i-\epsilon_j}
 \big<\phi_i \big| \hat{m} \big| \phi_j \big> \big<\phi_j \big| \hat{m} \big| \phi_i \big> \nonumber \\
 &=& 2 \f{T U^3}{B^4} \f{1}{\delta A}
 \label{eqn:chis}
\end{eqnarray}
and the density-density response for the 
interacting system
\begin{eqnarray}
 \chi &=& \sum_k \f{\left<\Psi_0 \left| \hat{m} \right| \Psi_k \right> \left<\Psi_k \left| \hat{m} \right| \Psi_0 \right>}{E_0-E_k} + c.c.\nonumber \\
 &=& 2 \f{U-B}{B(B+U)}.
 \label{eqn:chi}
\end{eqnarray}
For the two-site Hubbard model, these response functions are
just constants. The KS response has a functional dependence 
on the 1-matrix.  It diverges as the ground state is approached, 
i.e., in the limit $\delta A \rightarrow 0$. 
The linearized iteration map (\ref{eqn:map:lin}) is simply 
multiplication by a constant 
\begin{equation}
 \chi_s \chi^{-1} = -\f{U^2 (U+B)^2}{T B^3} \f{1}{\delta A} = \xi,
 \label{eqn:threshold:exp}
\end{equation}
which agrees with the direct calculation (\ref{eqn:xi:val}).  

\begin{figure}[ht!]
\includegraphics[width=0.9\columnwidth]{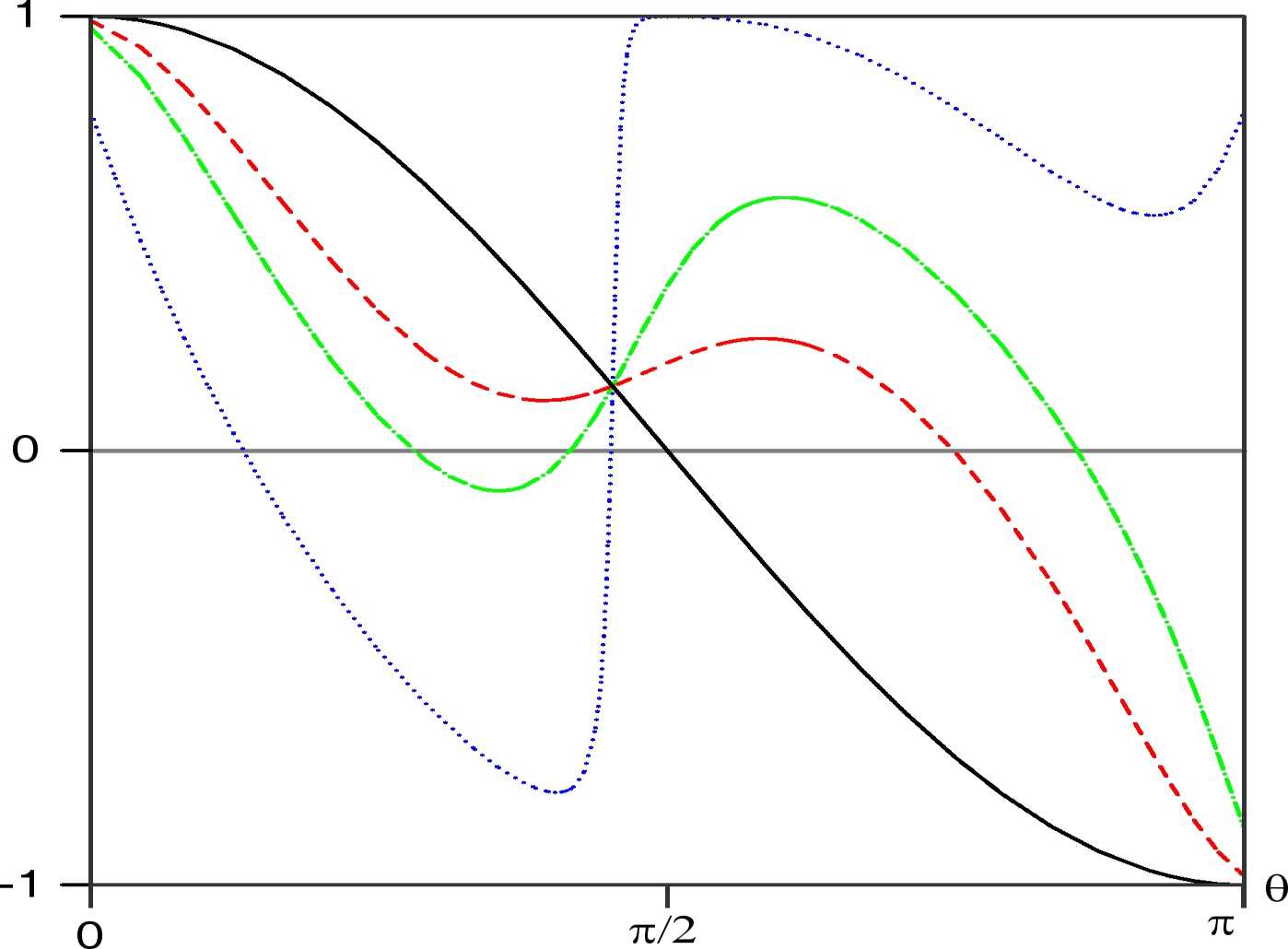}
\caption{\label{fig:shifting} [color online] The iteration map for $t=1$, $U=5$,
$V=-2.5$, and $A=A_{gs}-0.1$.  The solid [black] curve is the left hand side 
of (\ref{eqn:map}). The (dotted [blue], dash-dotted [green] , dashed [red]) curve is the right
hand side with level shift $\zeta=(0,3,6)$.  The threshold level
shift for convergence is $\zeta_c \approx 4.07$, which may be 
calculated with (\ref{eqn:criterion}).}
\end{figure}

Of course, in actual calculations it is necessary to have a convergent iteration
scheme. One possibility for obtaining convergence is the level shifting 
method,\cite{saunders1973} whose application in 1MFT was discussed in 
Sec.~\ref{ssec:shifting}.  In the level shifting method, one 
introduces artificial shifts of the KS eigenvalues in order to 
improve convergence.  A shift of the KS eigenvalue $\epsilon_i$ 
by an amount $\zeta_i$ is equivalent to adding a term 
$\zeta_i \big| \phi_i \big> \big< \phi_i \big|$ to 
the KS Hamiltonian, where $\phi_i$ is the orbital with
eigenvalue $\epsilon_i$.  The KS system for the two-site Hubbard
model has two orbitals.  As the divergence of the iteration
map is due to the degeneracy of the KS spectrum at the ground
state, it seems sensible to prevent degeneracy by introducing
a separation $2\zeta$ between the levels.  Thus, we add the 
following term to the KS Hamiltonian at iteration step $n$
\begin{equation}
 -\zeta \big| \phi_a \big> \big< \phi_a \big| 
 + \zeta \big| \phi_b \big> \big< \phi_b \big|
 = -\zeta \LB \sin \theta_n \s_x + \cos \theta_n \s_z \RB,
\label{eqn:shift}
\end{equation}
where $\phi_a$ and $\phi_b$ are evaluated at $\theta=\theta_n$.  
An example of the effect of level shifting is shown in Fig. \ref{fig:shifting}.
Convergence is achieved when $\zeta$ exceeds a threshold, which 
may be calculated from the convergence criterion (\ref{eqn:criterion}).  
The dashed [red] curve in Fig. \ref{fig:shifting} 
shows the iteration map with a level shift value greater than the threshold.  
For the two-site Hubbard model, the criterion for convergence can be 
visualized graphically as the condition that the magnitude of the slope 
of the level shifted curve be less than the slope of the solid [black] curve at 
the fixed point.

\begin{figure}[ht!]
\includegraphics[width=0.9\columnwidth]{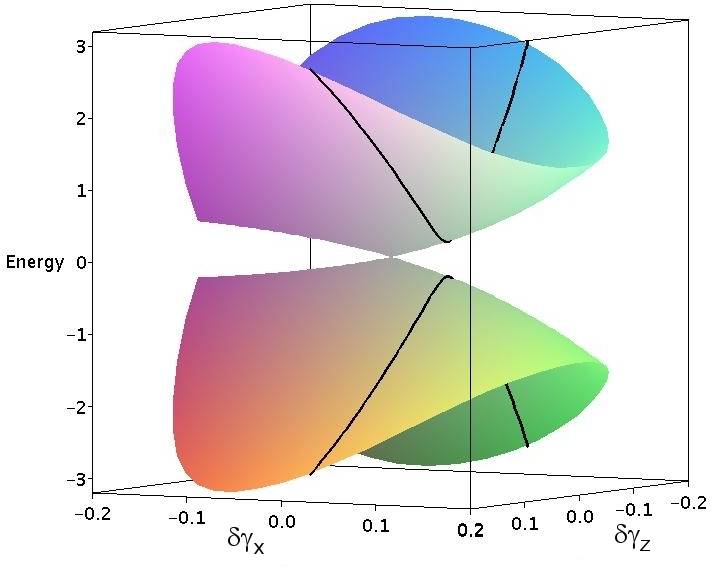}
\caption{\label{fig:curve} [color online] The KS energy for $t=1$, $U=3.5$ and $V=0$ 
is shown as a function of the deviation 
$(\delta \gamma_x, \delta \gamma_z)$ from the ground state.  
The two surfaces represent the two branches of the KS energy.  
The space curves show the energy as a function of $\theta$ for 
fixed occupation numbers. 
Optimization of the orbitals corresponds to moving along 
one of these curves to the stationary point. The ground 
state is the point of conic intersection at the origin.}
\end{figure}

At each iteration step the KS system has an ``instantaneous''
energy $E_s[\gamma] = tr ( \hat{h}[\gamma] \gamma )$, which 
has, of course, no physical meaning when the KS system is not
self-consistent.  The KS energy is shown
in Fig.~\ref{fig:curve} as a function of the deviation 
$\delta \vec{\gamma} = (\delta \gamma_x, \delta \gamma_z)$ of the 
1-matrix (\ref{eqn:1-matrix:param}) from the ground state 1-matrix.  
It is immediately seen that the KS energy is not stationary
at the ground state 1-matrix, which is a cusp point where the 
energy $E_s$ changes linearly with respect to $\delta \gamma$. 
The KS energy is multivalued due to the choice implied in 
occupying the KS levels when the system is not self-consistent 
(see Sec.~\ref{ssec:props}). The space curve in Fig.~\ref{fig:curve} 
shows the energy as a function of $\theta$ for fixed occupation 
numbers, i.e., for fixed $A$.  The KS response is proportional 
to the inverse separation between the two 
branches of the space curve.  The separation vanishes as the
curve approaches the conic point, which is the origin of the 
divergent KS response.  

\subsection{\label{ssec:DFT} Solution by DFT}

The two-site Hubbard model with the local external potential
chosen in (\ref{eqn:Hamiltonian2}) may be treated also with DFT.  
It is interesting to compare the DFT-KS scheme with the 1MFT-KS
scheme, especially with regard to their convergence behavior. 
The variational energy functional and KS Hamiltonian may be 
constructed explicitly. An interesting result of the investigation 
is that the straightforward iteration map is divergent when 
$U > 1.307 t$ (for $V=0$).  We derive a general condition for the 
convergence of the DFT-KS equations.

\subsubsection{\label{sssec:DFT:E} Energy functional}

The HK energy functional for a lattice is
\begin{equation}
 E[n,v]= \sum_i v(i) n_i + F[n],
 \label{eqn:E:HK}
\end{equation}
where $v(i)$ is the external potential at site $i$ and $F[n]$ is a 
universal functional of the density (here, site occupancy) defined as
\begin{equation}
 F[n] = \big< \Psi_0[n] \big| \hat{T}+\hat{U} \big| \Psi_0[n]\big>,
 \label{eqn:F}
\end{equation}
where $\hat{T}$ is the kinetic energy operator and $\hat{U}$ 
is electron-electron interaction. In the following treatment of the 
two-site Hubbard model, we depart from standard practice by enforcing 
the normalization condition $n_1+n_2=N$ explicitly (i.e., through 
the parametrization), rather than with a Lagrange multiplier.  
Thus, we take as basic variable the single parameter 
$m=(n_1-n_2)/2$ that uniquely specifies the density (site occupancy).  
Similarly, the external potential is specified by the single 
parameter $V=v(1)-v(2)$.  The functional $F[n]$ in (\ref{eqn:F}) is then 
just a function $F(m)$, which may be constructed explicitly as follows:  
i) a map $m \rightarrow \Psi_0$ is defined as the 
composition of the maps $m \rightarrow V$ and 
$V \rightarrow \Psi_0$ and ii) the resulting function $\Psi_0(m)$
is used to evaluate (\ref{eqn:F}). 
An explicit expression for the map $m \rightarrow V$ can be 
found from the inverse of (\ref{eqn:mV}). The second map $v \rightarrow \Psi_0$ was 
given in (\ref{eqn:Psi}).  The composition of these two maps
gives the ground state as a function of $m$, i.e.,
$\Psi_0(m)$, with which the universal functional (\ref{eqn:F})
may be evaluated.

\subsubsection{\label{sssec:DFT:h} KS Hamiltonian}

Following standard practice, the KS Hamiltonian takes
over, unchanged, the kinetic energy operator from
the many-body Hamiltonian.  Thus, we consider the
KS Hamiltonian
\begin{equation}
 \hat{h} = \hat{t} + \hat{v}_s,
 \label{eqn:DFT:h}
\end{equation}
where $\hat{t}=-t(c_1^{\dag} c_2 + c_2^{\dag} c_1)$
is the kinetic energy operator and $v_s(i)$ is the 
KS potential at site $i$ defined by
\begin{equation}
 v_s(i) = \f{\partial W}{\partial n_i},
 \label{eqn:vs}
\end{equation}
where $W[n]=E[n,v]-T_s[n]$ contains the Hartree and
exchange-correlation energy as well as the external 
potential energy, and $T_s[n]$ is the kinetic energy 
of the KS system. We do not separate these contributions 
explicitly.  The KS potential is spin independent 
because the ground state density is spin unpolarized. 
Also, it is determined only to within an arbitrary 
additive constant, which we choose such that 
$v_s(1)+v_s(2) = 0$.  In the site basis, the KS 
Hamiltonian is a $2\times 2$ matrix which may be 
expressed as $h = -t \sigma_x + (V_s/2) \sigma_z$,
where $V_s = v_s(1)-v_s(2)$.
The kinetic energy of the KS system is evaluated as
\begin{eqnarray}
 T_s &=& \sum_i^{occ} f_i \big< \phi_i \big| \hat{t} \big| \phi_i \big> \nonumber \\
 &=& 2 \big< \phi_a \big| \hat{t} \big| \phi_a \big> \nonumber \\
 &=& -2t \sin \theta,
 \label{eqn:Ts}
\end{eqnarray}
where $\phi_a$ is the lowest energy eigenstate of 
(\ref{eqn:DFT:h}) and is twice occupied (once by a spin
up electron and once by a spin down electron.)  It is
parametrized as in (\ref{eqn:orbitals}) with $\tan \theta = -2t/V_s$.  The density
of the KS system is
\begin{eqnarray}
 m_s &=& \sum_i^{occ} f_i \big< \phi_i \big| \hat{m} \big| \phi_i \big> \nonumber \\
 &=& \cos \theta.
 \label{eqn:ms}
\end{eqnarray}
Thus, from (\ref{eqn:Ts}) and (\ref{eqn:ms}) the kinetic energy $T_s$ is 
a known function of $m_s$. 
From (\ref{eqn:E:HK}), (\ref{eqn:vs}) and (\ref{eqn:Ts}),
the KS potential is
\begin{eqnarray}
 V_s &=& \left. \f{\partial W}{\partial m}\right|_{m=m_s} \nonumber \\
 &=& \left. \f{\partial}{\partial m} \LB E(m,V) - T_s(m) \RB \right|_{m=m_s} \nonumber \\
 &=& V + f(m_s) - \left.\f{\partial T_s}{\partial m}\right|_{m=m_s}
 \label{eqn:Vs}
\end{eqnarray}
where $V=v(1)-v(2)$ is the given external potential
and 
\begin{equation}
 f(m_s) = \left. \f{\partial F}{\partial m} \right|_{m=m_s}.
\end{equation}
Eq.~\ref{eqn:Vs} is simply the familiar expression 
$v_s(r) = v(r) + v_H(r) + v_{xc}(r)$ with a different partitioning
of the terms.
It is seen that the terms $f-\partial T_s/\partial m$ together 
correspond to the Hartree and exchange-correlation potentials.

\subsubsection{\label{sssec:iter} Iteration of the KS equations}

Let us investigate the iteration of the KS equations in the
present context.  The conventional iteration map consists of 
the following steps:
i) the KS potential for step $n+1$ is determined from the 
density of step $n$ using (\ref{eqn:Vs}), i.e., 
$V_s^{(n+1)} = V_s(m_s^{(n)})$,  
ii) the eigenstates of $\hat{h}^{(n+1)}$ are found, and
iii) the density of step $n+1$ is calculated with (\ref{eqn:ms}).

Consider step (i) in more detail.  The KS potential is 
obtained from (\ref{eqn:Vs}),
\begin{equation}
 V_s^{(n+1)} = V + f(m_s^{(n)}) - \left.\f{\partial T_s}{\partial m}\right|_{m=m_s^{(n)}}.
 \label{eqn:Vs2}
\end{equation}
The right hand side may be expressed differently by using the
stationary conditions for the energy functional $E[n,v]$ and
the KS energy $E_s=T_s+\sum_i v_s(i) n_i$. 
The stationary condition $\partial E/\partial m = 0$ applied 
to (\ref{eqn:E:HK}), gives $f=-V(m)$, where $V(m)$ 
is the external potential such that the interacting system 
has ground state density $m$.  
Similarly, the stationary condition applied to $E_s$ gives
$\partial T_s/\partial m = -V_s$.
Substituting these relations in (\ref{eqn:Vs2}) yields
\begin{equation}
 V_s^{(n+1)} = V - V(m_s^{(n)}) + V_s(m_s^{(n)}).
 \label{eqn:Vs3}
\end{equation}
At self-consistency the $V_s$ terms cancel, and we obtain
the expected result $V=V(m_{gs})$, where $m_{gs}$ is the 
ground state density.  
For the present model, the ground state density could be
found by solving $V=V(m)$ as $V(m)$ is known exactly from (\ref{eqn:mV}). 
However, in general the ground state must be found by iteration.  
Eq.~\ref{eqn:Vs3} implies an iteration map for the density, 
i.e., a map 
$m_s^{(n)} \rightarrow m_s^{(n+1)}$, because $V_s^{(n+1)}$
determines $m_s^{(n+1)}$.  From (\ref{eqn:ms}) and the 
definition $\tan \theta = -2t/V_s$, we find the 
relationship
\begin{equation}
 V_s = -2t \f{m_s}{\sqrt{1-m_s^2}}.
 \label{eqn:Vsm}
\end{equation}

The density may be iterated until self-consistency is 
reached.  However, we encounter a technical difficulty
for the present model.  In order to express explicitly 
the term $V(m_s^{(n)})$ in (\ref{eqn:Vs3}), we must 
invert (\ref{eqn:mV}), which involves solving a cubic
equation.  As the solutions are rather unwieldy, we 
take here a different approach.  We iterate 
instead the external potential $V(m)$.  It may seem odd 
to iterate the external potential, which is given in
the statement of the problem.  Nevertheless, the 
iteration map for $V$ provides an ``image'' of the
iteration map for $m_s$, by virtue of the HK theorem.
Such an approach allows us to investigate certain
features of the iteration map, in particular
its convergence behavior.   
In order to express (\ref{eqn:Vs3})
as an iteration map for $V$, we need to express $V_s$ as a
function of $V$.
In other words, we find the value of $V_s$ such that
the KS system has density $m_s=m$, where $m$ is the
density of the interacting system with $V$. The 
composition of (\ref{eqn:Vsm}) and (\ref{eqn:mV}) 
yields the desired function 
\begin{equation}
 \tilde{V}_s(V)=V_s(m(V)).
 \label{eqn:comp}
\end{equation}
Using (\ref{eqn:comp}) in (\ref{eqn:Vs3}), we obtain
the iteration map for the external potential 
\begin{equation}
 \tilde{V}_s(V^{(n+1)}) = V - V^{(n)} + \tilde{V}_s(V^{(n)}),
 \label{eqn:Vmap}
\end{equation}
which is expressed in implicit form. 

\begin{figure}[ht!]
\centering
\includegraphics[width=0.9\columnwidth]{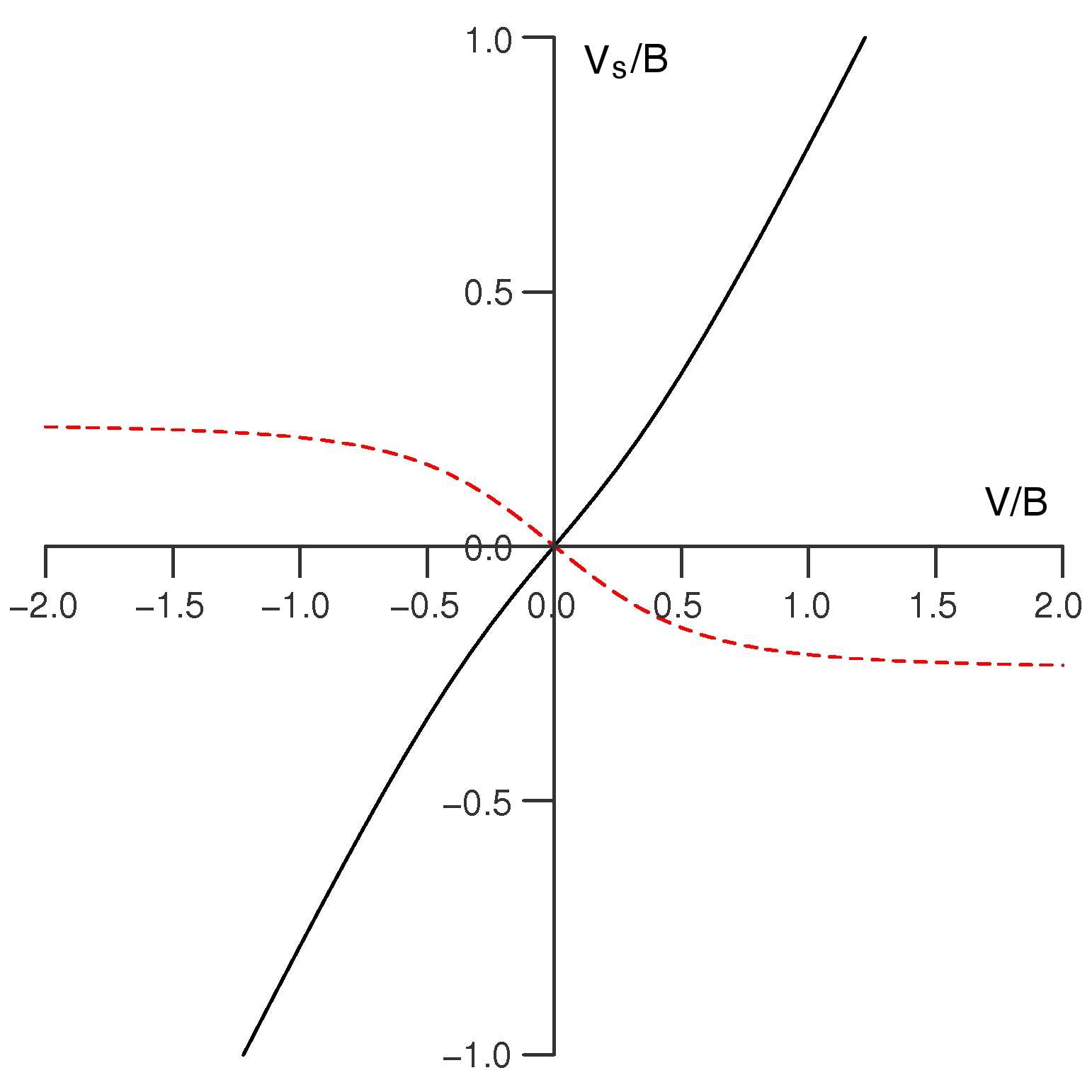}
\caption{\label{fig:Vmap:weak} The iteration map
for the external potential $V$ is shown for a  
weakly interacting system with $U=t$. The left and 
right hand sides of (\ref{eqn:Vmap}) are shown as solid [black] 
and dashed [red] curves, respectively.}
\end{figure}

\begin{figure}[ht!]
\centering
\includegraphics[width=0.9\columnwidth]{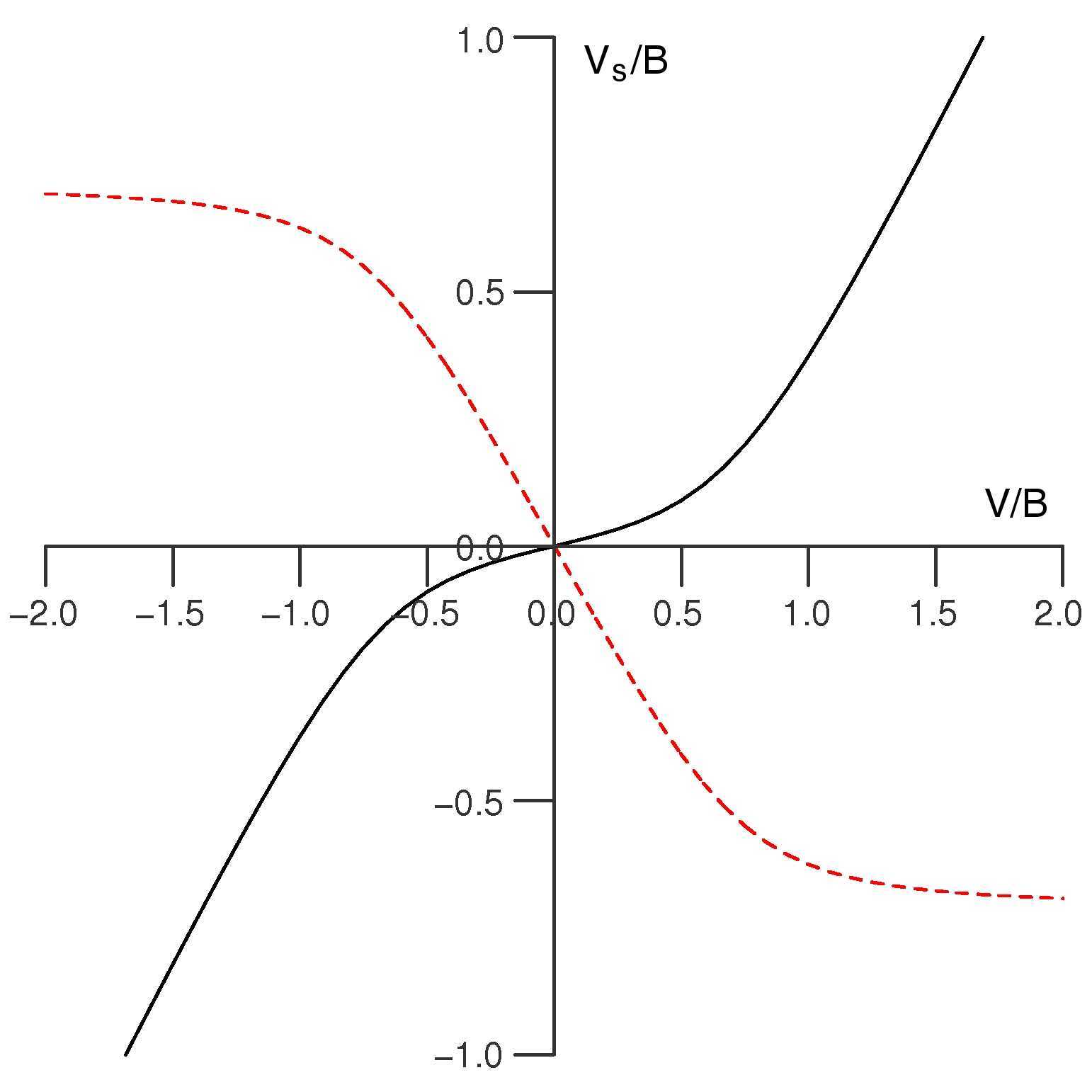}
\caption{\label{fig:Vmap:strong} The iteration map
for the external potential $V$ is shown for a  
strongly interacting system with $U=4t$. The left and 
right hand sides of (\ref{eqn:Vmap}) are shown as solid [black] 
and dashed [red] curves, respectively.}
\end{figure}

Examples of the iteration map for a uniform system ($V=0$) 
are shown in Figs. \ref{fig:Vmap:weak} and \ref{fig:Vmap:strong}, 
where the left and right hand sides of (\ref{eqn:Vmap})
are plotted.
Suppose an initial value $V^{(0)}\neq0$ is chosen.  
For a system with $V=0$, the ground state has uniform density ($m=0$),
but the initial density $m_s^{(0)}$ associated with $V^{(0)}$ is not 
uniform.  Upon iteration, we expect the KS system to relax 
to a uniform density, i.e., we expect the KS potential to be 
such as to push the system closer to uniform occupancy in 
the next iteration. 
The solid [black] curves in Figs.~\ref{fig:Vmap:weak} and 
\ref{fig:Vmap:strong} represent the left hand side of (\ref{eqn:Vmap}),
while the dashed [red] curves represent the right hand side.
The iteration map may be demonstrated graphically by alternately drawing
vertical lines from the solid curve to the dashed curve and 
horizontal lines from the dashed curve to the solid curve. The map displays
``charge oscillation.'' 
The ground state is a stable fixed point if the magnitude of the 
slope of the dashed curve at the origin is less than the slope of the 
solid curve at the origin.  For weakly interacting
systems the iteration map is convergent, while for strongly 
interacting systems it is nonconvergent. 
The threshold for convergence is $U\approx 1.307 t$.  

\subsubsection{\label{sssec:lin} Linearization of the KS equations}

The nature of the fixed point and the origin of diverent 
behavior are revealed by linearization of the
iteration map.  We linearize the map  
by expanding both sides of (\ref{eqn:Vs3}) with respect
to $\delta m_s = m_s-m_{gs}$, where $m_{gs}$ is the ground state density.  
We find
\begin{eqnarray}
 \chi_s^{-1} \delta m_s^{(n+1)} &\approx &\chi_s^{-1} \delta m_s^{(n)} - \chi^{-1} \delta m_s^{(n)} \nonumber \\
 \delta m^{(n+1)} &\approx &  \chi_s \LB \chi_s^{-1} - \chi^{-1} \RB \delta m_s^{(n)},
 \label{eqn:mapVs:lin}
\end{eqnarray}
where $\chi_s$ and $\chi$ are the density-density response
functions defined in (\ref{eqn:chis}) and (\ref{eqn:chi}).
The threshold for convergent behavior is
\begin{equation}
 \left| 1 - \chi_s \chi^{-1} \right| \leq 1,
 \label{eqn:threshold}
\end{equation}
or equivalently, $\chi_s \chi^{-1} \leq 2$.  Note the change 
from 1 for 1MFT to 2 for DFT on the right hand side, cf.  
(\ref{eqn:map:lin}).
Consider the case $V=0$, which has 
uniform density in the ground state ($m_{gs}=0$).  
Using the (\ref{eqn:chis}) and (\ref{eqn:chi}) in (\ref{eqn:threshold}), 
gives the threshold condition
\begin{equation}
 \cos(\pi/4-\alpha_0/2) = 4 \LB \sin(\pi/4-\alpha_0/2) \RB^3.
\end{equation}
The threshold is $\cos(\pi/4-\alpha_0/2) \approx 0.8095$, which 
corresponds to $U \approx 1.307 t$.  Let us consider
the limit $U \rightarrow \infty$.  The leading behavior
of the KS response is independent of $U$,
\begin{equation}
 \chi_s \sim \f{1}{T},
\end{equation}
while the response of the interacting system vanishes as
\begin{equation}
 \chi \sim \f{1}{4} \f{T^2}{U^3}.
\end{equation}
For sufficiently large $U$, the threshold (\ref{eqn:threshold})
is crossed and divergent behavior results.  
In DFT, as also in 1MFT, the source of divergent behavior is a 
KS response that is too large in relation to the exact response. 
In 1MFT the imbalance results from a divergent KS response, 
whereas in DFT the KS response generally remains finite but the
response of the interacting system becomes too small as $U$ 
increases.

In standard DFT (with continuous $n(r)$), the analog of the 
linearized iteration map (\ref{eqn:mapVs:lin}) may be written
\begin{eqnarray}
 n^{(n+1)}(r) &\approx& \int dr\pr dr^{\prime \prime} \chi_s(r,r\pr) 
\LB v_c(r\pr,r^{\prime\prime}) + f_{xc}(r\pr,r^{\prime\prime}) \RB \nonumber\\ 
 &&\times n^{(n)}(r^{\prime\prime}), 
\end{eqnarray}
where $n^{(n)}(r)$ is the density of iteration step $n$, 
$v_c$ is the kernel of the Coulomb interaction,
and $f_{xc} = \delta v_{xc}/\delta n$ is the 
exchange-correlation kernel.  The necessary and 
sufficient condition for convergence of the KS equations
is that all eigenvalues of the operator
\begin{equation}
 \hat{\chi}_s \big( \hat{v}_c + \hat{f}_{xc} \big)
\end{equation}
have modulus less than 1.

\section{Conclusions}

The status of the KS system in 1MFT has been uncertain.
Although Gilbert derived effective single-particle 
equations from the stationary conditions for the energy 
functional, the degeneracy of essentially all of the 
resulting orbitals was thought to be 
paradoxical. \cite{gilbert1975,tal1985,valone1980}
We have presented an alternative derivation of the KS 
equations in which the degeneracy is lifted by 
constraining the occupation numbers.  Such a KS scheme is 
well-behaved in the neighborhood of the ground state 
occupation numbers.  Therefore, the correct natural orbitals 
are obtained in the limit that the ground state is approached.   
We have constructed explicitly the 1MFT-KS system for a
simple two-site Hubbard model.  While we find no paradoxical 
results, the KS system has many striking features, in 
particular the collapse of eigenvalues at the ground state.
Although the KS eigenvalues do not have a physical 
interpretation as in DFT, the orbitals, which are 
called natural orbitals, play an important role in the
context of configuration interaction, i.e., the 
expansion of the full wavefunction as a sum of Slater
determinants. \cite{loewdin1955}  This may be important
in the search for approximate energy functionals.

Beyond the question of the existence of the KS system
in 1MFT, there is the issue of its practicality. 
The KS system has been extremely useful in DFT calculations.  
Due to the implicit 1-matrix dependence of the single-particle 
potential, the KS equations are nonlinear.  Such equations are 
generally solved by iteration.  As in DFT, there is a
``straightforward'' procedure for iteration.  In contrast to DFT, 
the ``straightforward'' procedure is always divergent,
in the sense that the ground state is an unstable fixed point.
We have demonstrated the instability of the ground state by 
linearization of the iteration map.  The source of the 
instability is the divergence of the KS static response 
function at the ground state, which in turn, is due to the
degeneracy of the KS spectrum.  Degeneracy-driven instability
is reminiscent of the Jahn-Teller effect, and the connection
is strengthened if we regard the implicit 1-matrix dependence
of the KS Hamiltonian as analogous to the parametric dependence
of the Born-Oppenheimer Hamiltonian on nuclear coordinates.  
In both cases, the energy spectrum splits linearly with respect 
to displacement from the degeneracy point.  Thus, the energy
may always be lowered by displacement.  For the 1MFT-KS system,
this means that the KS energy $tr(\hat{h} \hat{\ga})$ may always be 
lowered by displacement from the ground state, leading to an 
instability of the iteration procedure.  However, this is a 
fictitious energy and the HK energy functional $E_v$ is
of course always minimum at the ground state.

\begin{acknowledgments}
We gratefully acknowledge helpful discussions with Wei Ku. 
\end{acknowledgments}

\appendix

\section{Ground state not determined by the density}
We give here a simple example which shows that the density alone does not 
always uniquely determine the ground state wavefunction if the external 
potential is nonlocal.  Our example is the two-site Hubbard model,
which was solved for the case of a local external potential in 
Sec.~\ref{ssec:direct}.  The Hamiltonian $\hat{H}=\hat{T}+\hat{V}+\hat{U}$ 
is given in (\ref{eqn:Hamiltonian2}).  In such a lattice model, the 
hopping parameters $t_{jk}$ are real numbers that represent the kinetic 
energy. A ``magnetic field'' can be introduced by giving $t_{jk}$ a phase, 
i.e., by the transformation 
$t_{jk} \rightarrow t_{jk} \textrm{exp}\big(i \sum_{n=j}^k A(n)\big)$,
where $A(n)$ is the ``vector potential'' at site $n$ and the sum runs 
over a string of sites from site $j$ to site $k$. For the two-site model 
this is just the transformation $t \rightarrow t e^{i \tau}$. 
We see that this magnetic field appears in the Hamiltonian in 
exactly the same manner as a nonlocal external potential, such as
$v_{12} c_1^{\dag}c_2 + h.c.$, because it modifies the nonlocal hopping 
terms.  We can generate the above phase transformation
by the $U(1)$ rotations $c_1 \rightarrow c_1 e^{i\tau/2}$ and 
$c_2 \rightarrow c_2 e^{-i\tau/2}$.  The eigenstates of the transformed 
Hamiltonian are readily generated from the eigenstates of the original 
Hamiltonian by applying the same transformation. 
For example, without the magnetic field, the ground state to first order 
in small $V$ is 
\begin{equation}
 \Psi_0 = \Phi_0 + \f{\big<\Phi_2\big| \hat{V} \big| \Phi_0\big>}{E_0-E_2} \Phi_2,
\end{equation}
where $\Phi_i$ are given in (\ref{eqn:Phi}).  When the magnetic 
field is turned on, the $\Phi_i$ change, e.g.,
\begin{equation}
 \Phi_0 \rightarrow \f{1}{\sqrt{2}} \LB \begin{array}{c} y e^{-i\tau} \\ x \\ x \\ y e^{i\tau} \end{array} \RB
\end{equation}
in the site basis 
$\big\{c_{1\u}^{\dag} c_{1\d}^{\dag} \left| 0\right>$, 
$c_{1\u}^{\dag} c_{2\d}^{\dag} \left| 0\right>$, 
$c_{2\u}^{\dag} c_{1\d}^{\dag} \left| 0\right>$,
$c_{2\u}^{\dag} c_{2\d}^{\dag} \left| 0\right>\big\}$.
Accordingly, the ground state acquires a nontrivial dependence on the
magnetic field ($\tau$-dependence).  At the same time, the ground state 
1-matrix is transformed to 
\begin{equation}
 \ga = \ga_0 + A_0 \LB \begin{array}{lr} \cos \theta_0 & \sin \theta_0 e^{-i\tau} \\
 \sin \theta_0 e^{i\tau} & -\cos \theta_0
\end{array} \RB,
\end{equation}
where $A_0$ and $\theta_0$ are the ground state values (for $\tau=0$) 
of the variables defined in (\ref{eqn:orbitals}) and (\ref{eqn:1-matrix:param}).
The density is given by the diagonal elements, which are  
unaffected by the transformation.  Only the off-diagonal (nonlocal)
elements are sensitive to the magnetic field.  Therefore, the
1-matrix rather than the density is required to uniquely specify 
the ground state.\cite{gilbert1975}


\bibliography{article}

\end{document}